\documentclass[]{aastex631}

\usepackage{subfigure}
\usepackage{longtable}
\usepackage{threeparttable}

\begin{document}

\title{Evidence for a close-in tertiary orbiting around the Algol-type system HZ Dra with tidal splitting and spots activities}

\correspondingauthor{Wen-Ping Liao}
\email{$^{*}$liaowp@ynao.ac.cn}	
\author{Ping Li}
\affiliation{Yunnan Observatories, Chinese Academy of Sciences (CAS), 650216 Kunming, China}
\affiliation{University of Chinese Academy of Sciences, No.1 Yanqihu East Rd, Huairou District, Beijing, China 101408}

\author{Wen-Ping Liao$^{*}$}
\affiliation{Yunnan Observatories, Chinese Academy of Sciences (CAS), 650216 Kunming, China}
\affiliation{University of Chinese Academy of Sciences, No.1 Yanqihu East Rd, Huairou District, Beijing, China 101408}

\author{Sheng-Bang Qian}
\affiliation{School of Physics and Astronomy, Yunan University, Kunming 650091, China}

\author{Lin-Jia Li}
\affiliation{Yunnan Observatories, Chinese Academy of Sciences (CAS), 650216 Kunming, China}

\author{Qi-Bin Sun}
\affiliation{Yunnan Observatories, Chinese Academy of Sciences (CAS), 650216 Kunming, China}
\affiliation{University of Chinese Academy of Sciences, No.1 Yanqihu East Rd, Huairou District, Beijing, China 101408}

\author{Xiang-Dong Shi}
\affiliation{Yunnan Observatories, Chinese Academy of Sciences (CAS), 650216 Kunming, China}

\author{Azizbek Matekov}
\affiliation{Yunnan Observatories, Chinese Academy of Sciences (CAS), 650216 Kunming, China}
\affiliation{University of Chinese Academy of Sciences, No.1 Yanqihu East Rd, Huairou District, Beijing, China 101408}

\author{Qi-Huan Zeng}
\affiliation{Yunnan Observatories, Chinese Academy of Sciences (CAS), 650216 Kunming, China}
\affiliation{University of Chinese Academy of Sciences, No.1 Yanqihu East Rd, Huairou District, Beijing, China 101408}

\author{Zhao-Long Deng}
\affiliation{Yunnan Observatories, Chinese Academy of Sciences (CAS), 650216 Kunming, China}
\affiliation{University of Chinese Academy of Sciences, No.1 Yanqihu East Rd, Huairou District, Beijing, China 101408}

\author{Xiao-Hui Fang}
\affiliation{School of Mathematics, Physics and Finance, Anhui Polytechnic University, Wuhu 241000, China}



\begin{abstract}

We reported a cyclic variation of $O-C$ diagram with a semi-amplitude of 0.0033 days and a period of 1.05 years for the pulsating eclipsing binary HZ Dra. The cyclic variation can be explained by the light travel-time effect via the presence of a close-in third body orbiting around HZ Dra in an elliptical orbit with a maximum semi-major axis of 0.92 au. Based on the W-D code, the contribution of the third light to the total system is determined to be 29 $\%$, which is in agreement with the estimated value. Our light curve modelling indicates an evolving hot and cool spot on the surface of the primary and secondary components, respectively. Their positions are roughly symmetrical to the inner Lagrangian point L1, which could be used to explain the variation in the O$^{'}$Connell effect. Our frequency analysis detects 1 radial p-mode, 7 non-radial p-modes and 1 non-radial g-mode. In addition, a total of 6 multiplets are identified, spaced by the orbital frequency, which can be explained as a tidally split mode caused by the equilibrium tides of the close binary system with a circular orbit. These pulsating features suggest that the primary of HZ Dra is a $\delta$ Scuti star, pulsating in both p- and g-mode and influenced by tidal forces.

\end{abstract}



\section{Introduction} \label{sec:intro}

$\delta$ Scuti stars are main sequence or slightly evolved post-main sequence stars with masses between 1.5 and 2.5 M$_{\odot}$.  Their pulsations are driven by the $\kappa$ mechanism involving He $_{\rm II}$ and H zones \citep{1999A&A...351..582H}. The mechanism causes typical oscillations in radial and non-radial p-modes (with pressure as the restoring force) and g-modes (with buoyancy as the restoring force) with a period range of about 0.02-0.30 days (e.g., \citealp{2010aste.book.....A,2018AJ....156...82C}). Many $\delta$ Scuti stars have been found to be a component of eclipsing binary systems. \cite{2018MNRAS.475..478Q} reported that 766 $\delta$ Scuti stars were observed by the Guoshoujing telescope (the Large-Sky Area Multi-Object Fibre Spectroscopic Telescope, LAMOST), where stellar atmospheric parameters of 525 variables were determined. Meanwhile, 88 $\delta$ Scuti stars were found to be candidates for binary or multiple systems. \cite{2022ApJS..259...50S} analyzed the light curves of 1626 EA-type binaries observed by the Transiting Exoplanet Survey Satellite (TESS) mission \citep{2015JATIS...1a4003R} and presented that 57 new $\delta$ Scuti stars in these type systems. \cite{2022ApJS..263...34C} then reported that 143 new eclipsing binaries containing $\delta$ Scuti stars were discovered using TESS data.

These targets, reported by \cite{,2022ApJS..259...50S} and \cite{2022ApJS..263...34C}, are Algol-type binary systems, where the primary component is an A-F type star and the secondary star is an F-K type giant or sub-giant, always accompanied by strong magnetic activity \citep{1989SSRv...50..205B}. According to the Algol paradox \citep{1998A&AT...15..357P}, the initially more massive star in the binary system evolves to fill its Roche lobe first and transfers mass to another component, causing to the mass ratio to reverse and the Algol-type system to form. Both single $\delta$ Scuti stars and those exist in eclipsing binaries show the same pulsation characteristics, but differ in their evolutionary process. The companion star in a binary system may have an effect on the pulsation components.  For example, in classical Algol-type binaries, a likely effect such as mass transfer, accretion, and gravitational force applied from the secondary star to the primary component can affect the periods, amplitudes, and modes of the pulsations \citep{2013AJ....145...87S}. If the massive component after mass transfer lies within the instability of $\delta$ Scuti, an oscillating eclipsing binary system of Algol-type (oEA; \citealp{2002ASPC..259...96M}) could be formed. In addition, the magnetic activity on the late-type secondary components can cause the orbital period of the Algol-type system to change \citep{1992ApJ...385..621A}. The possible third body around the oEA system, for example HL Dra \citep{2021MNRAS.505.6166S} and TZ Dra \citep{2022MNRAS.510.1413K}, can also cause the orbital period to vary \citep{1952ApJ...116..211I}. The existence and correlation of such phenomena make the investigation of oEA system is difficult and attractive.

 When $\delta$ Scuti stars are components of eclipsing binaries, the influence of tidal forces on the pulsator should be considered. \cite{2003A&A...404.1051R} studied the tidal effect on self-driven oscillations in one component of a binary rotating synchronously in a circular orbit. Their study showed that each (degenerate) pulsation eigenfrequency of a spherically symmetric star is then split into (2$l$+1) eigenfrequencies whose differences for a given $l$ have a regular spacing equal to (a multiple of) the orbital frequency.  The occurrence of tidal effects on stellar oscillations will be most visible in stars with large tidal distortions, which can occur when the orbital period is short and the stars are of comparable mass \citep{2018MNRAS.476.4840B}.
 
 HZ Draconis (HZ Dra) was discovered as an EA-type eclipsing binary star with a period of 0.7729340 days and an A0-type primary component by the Hipparcos mission \citep{1997yCat.1239....0E}. \cite{2012MNRAS.422.1250L} had observed the multi-colour light curves of HZ Dra and also obtained a single radial velocity curve of the primary component. Their results indicate that the eclipsing binary is an oEA system with a characteristic frequency of 52.17 d$^{-1}$ and a mass of 2.9 $(\pm0.1)$ M$_{\odot}$ for the primary. However, the $l$ degree of this frequency was not determined in their study. \cite{2017MNRAS.470..915K} had re-determined the spectral classification (A8 / A7V) for HZ Dra, which is significantly different from the previous A0 classification. Since then, no detailed studies of orbital period variations, pulsation features and spot activities for this system have been published. In addition, HZ Dra was observed by the TESS mission from 2019 to 2022, producing 31 sectors light curves (sectors: 14-26, 40-41, 47-53, 55-60, 73-75) with an exposure time of 120 seconds. The aim of this paper is to present a detailed study of orbital period variations, pulsation features and spot activities for HZ Dra mainly based on TESS data. \begin{figure*}[ht!]
 	\plotone{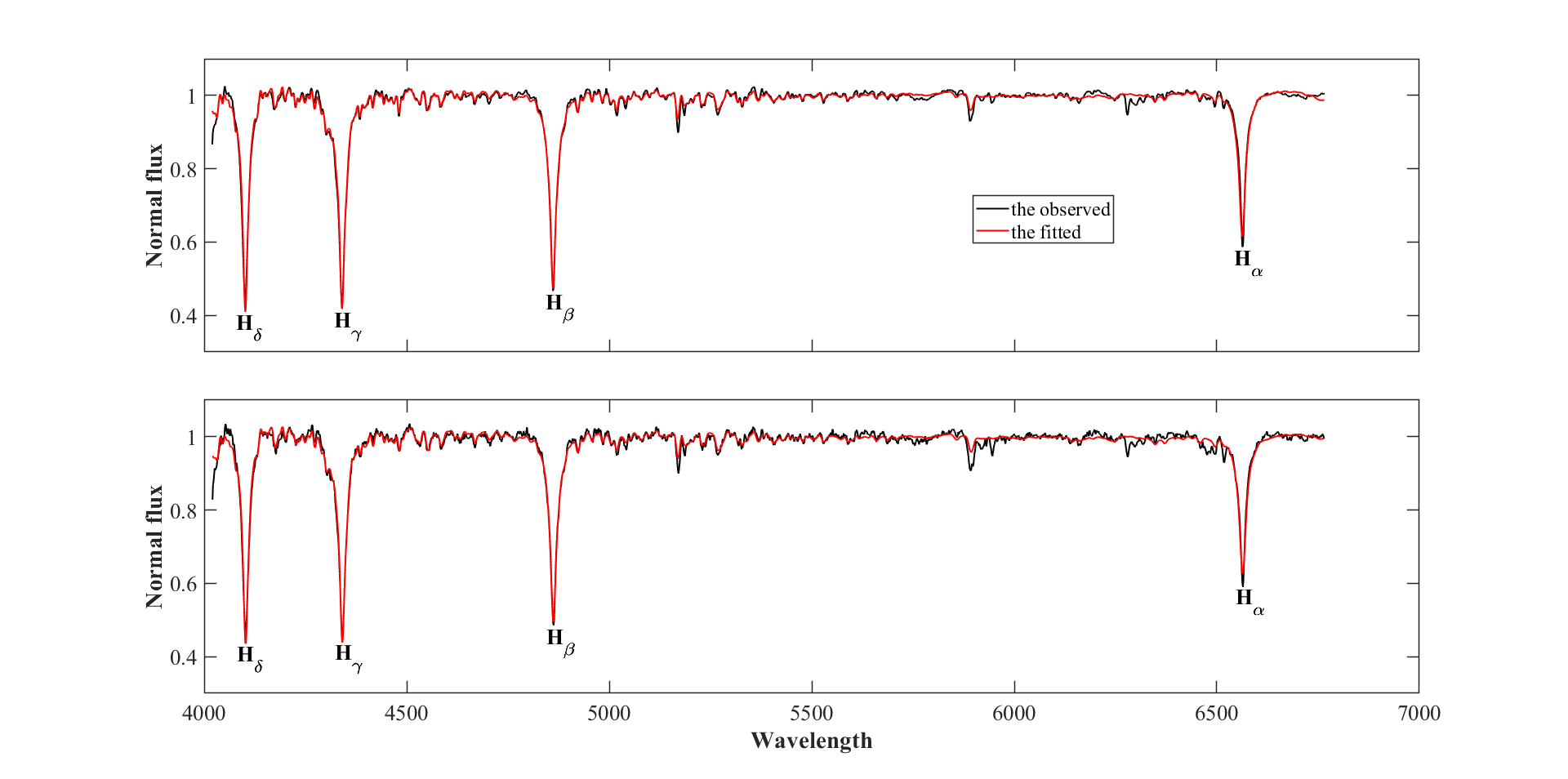}
 	\caption{Upper panel: spectrum 1 of HZ Dra observed on 2021 June 05. Bottom panel: spectrum 2 obtained on 2021 June 12. The black and red lines represent the observed and fitted spectra, respectively. The four Balmer's absorbed lines H$_{\delta}$, H$_{\gamma}$, H$_{\beta}$ and H$_{\alpha}$ are also marked.\label{fig:figure1}}
 \end{figure*}

\section{Spectroscopic Observations} \label{sec:Ob}
To determine the stellar atmospheric parameters of HZ Dra, spectra observations were performed on 05 June 2021 and 12 June 2021 with the Beijing Faint Object Spectrograph and Camera (BFOSC) mounted on the 2.16 m telescope at the Xinglong Station of the National Astronomical Observatories, Chinese Academy of Sciences. The BFOSC low-dispersion spectrometer and the G7 grism with a slit width of 1$^{''}$.8 were used, with a corresponding spectral resolution of about $\sim$ 700 and an observation wavelength range of 400-680 nm \citep{2016PASP..128k5005F}.

The two spectra with an exposure time of 600 s, shown in Figure \ref{fig:figure1}, were obtained after processing the observational data using IRAF software. The University of Lyon Spectroscopic Analysis Software (hereafter ULySS) \citep{ 2009A&A...501.1269K} was used to obtain stellar atmospheric parameters by fitting full spectra to mode spectra generated by an interpolator using the ELODIE library \citep{ 2001A&A...369.1048P}. Finally, the values of the stellar atmospheric parameters were derived from the two spectra observations and are presented in Table \ref{table 1}. The fitted spectra with four Balmer absorbed lines H$_{\delta}$, H$_{\gamma}$, H$_{\beta}$ and H$_{\alpha}$ are also shown with red lines in Figure \ref{fig:figure1}.
\begin{table*}
	\begin{center}
		\caption{Atmospheric parameters of the primary component of HZ Dra. The numbers in the brackets are only the fit errors provided by ULySS.}
		\label{table 1}
		\begin{tabular}{lccc}\hline
			Parameters &spectrum 1&spectrum 2& mean values\\\hline
			$[Fe/H]$ (dex)  & -0.27 (2) &-0.22 (3)&-0.24 (15) \\
			$\mathit{T}_{\rm eff}$ (K) & 8143 (19) & 8063  (21)&8103 (252)\\
			$\log g$ (cm s$^{\rm -2}$)& 3.99 (2) &3.93 (3)&3.96 (18) \\
			\hline
		\end{tabular}
	\end{center}
\end{table*}

In Table \ref{table 1}, the uncertainties of the atmospheric parameters provided by  ULySS are only the fit errors, and the real errors (external errors) of the atmospheric parameters may be about more 20 times larger than the fit errors \citep{ 2009A&A...501.1269K}. According to the photometric results of \cite{2012MNRAS.422.1250L}, the secondary star contributes about 0.8$\%$ luminosity to the total system in the V-band. Therefore, in the latter calculation process, we used the mean values of the two spectra as the atmospheric parameters of the primary component.

\section{Analysis for O$^{'}$Connell effect and Light curve model} \label{sec:WD}
\begin{figure*}[ht!]
	\plotone{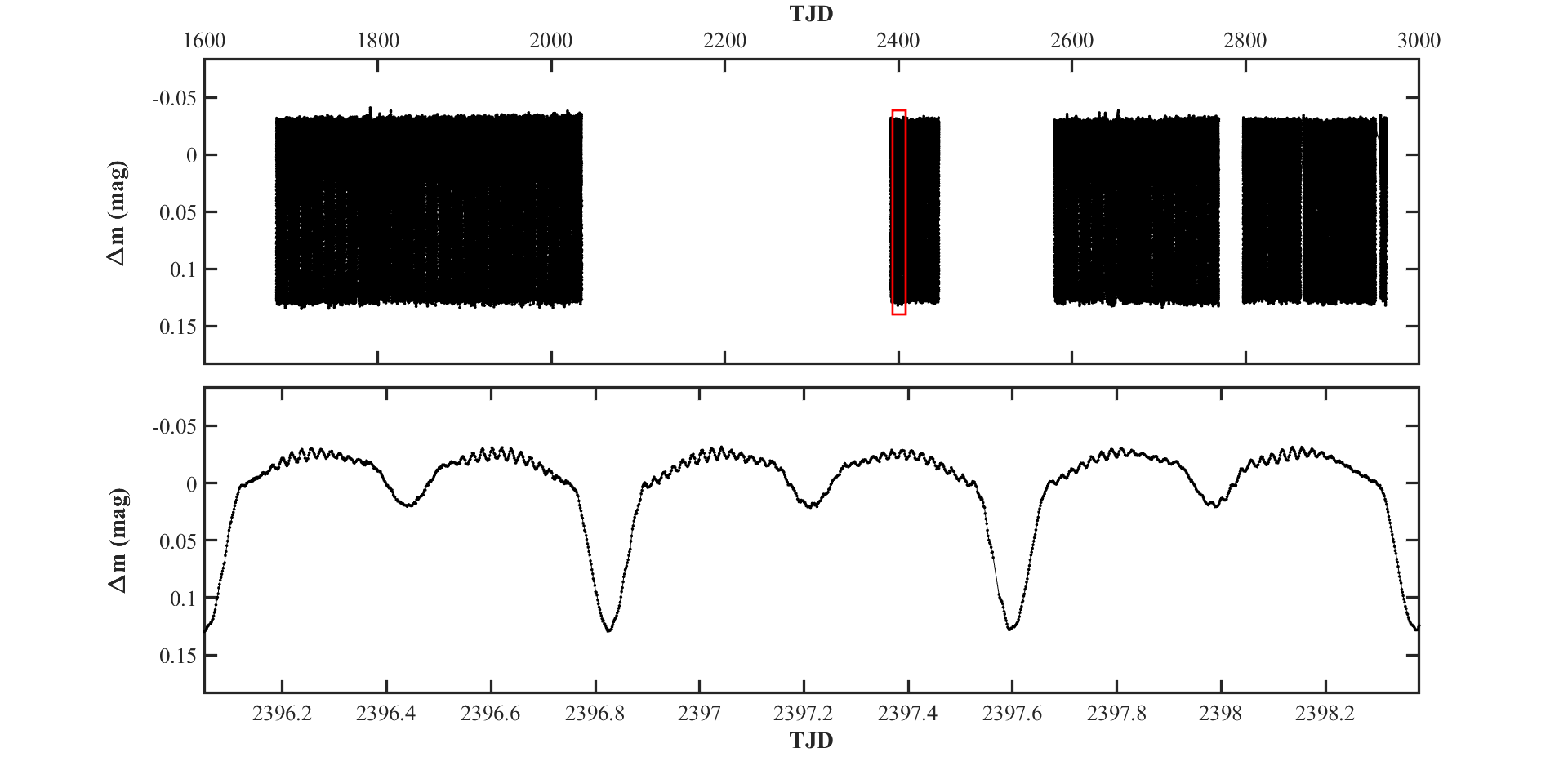}
	\caption{Upper panel: the TESS light curves of HZ Dra (TJD=BJD-2457000.0). Bottom panel: a section of the light curve is displayed in the enlarged panel.\label{fig:figure 2}}
\end{figure*}

The light curve data of HZ Dra observed with the TESS mission can be downloaded from the Mikulski Archive for Space Telescopes (MAST) \footnote{MAST: \href{https://archive.stsci.edu/mast.htm}{$archive.stsci.edu$}} database. The method described by \cite{2022ApJS..259...50S} was used to correct the flux. The corrected flux was then converted to magnitudes. Finally, all the TESS light curves for each sector of HZ Dra are shown in Figure \ref{fig:figure 2} against $\Delta$m and TJD (TJD=BJD-2457000.0), where the lower panel shows a section of the light curve in detail and the pulsations on the light curve are clearly visible. 

The O$^{'}$Connell effect is a phenomenon that the magnitude of the light maximum near 0.25 phase (hereafter Max I) differs from that near 0.75 phase (hereafter Max II; e.g., \citealp{1969CoKon..65..457M}; \citealp{1999oaaf.conf..377L}; \citealp{2014ApJS..212....4Q}), which can be caused by many common and complex mechanisms in eclipsing binaries. Such mechanisms include flares, spots, mass transfer, etc. Each of these factors can change the local brightness of the surface of the star in the binary system.

\begin{figure*}[ht!]
	\plotone{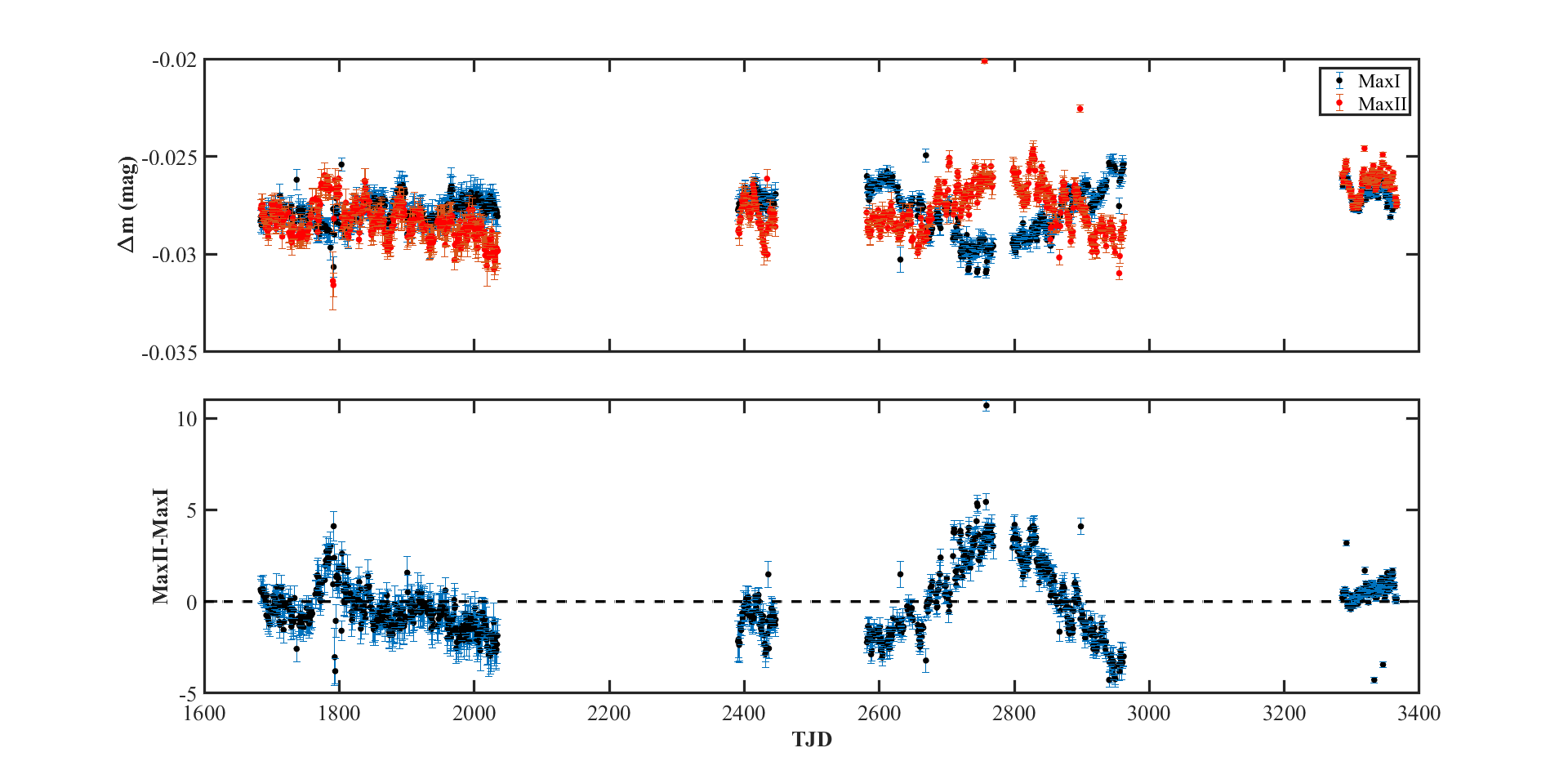}
	\caption{The variations of Max II, Max I and Max II-Max I.\label{fig:figure 3}}
\end{figure*}

To eliminate the possible effect of the O$^{'}$Connell effect, we firstly took into account of all the available TESS data to check whether it changes from one sector to another. We then fit the light curves and calculated the magnitudes of Max I, Max II using parabola fitting. As one can see from Figure \ref{fig:figure 3}, the variations of Max I and Max II are roughly in phase for a long time, but during TJD 1790-1810 and TJD 2600-3000 the variations of Max I and Max II cross each other and have a 180$^{\circ}$ phase difference.
According to \cite{2022MNRAS.510.1413K}, this phenomenon is due to the presence of spots whose position varies with time on the surface of the component(s). As shown in the lower panel of Figure \ref{fig:figure 3}, the Max I is equal to Max II near TJD 1700, which means that no significant variation was found between the levels of the out of eclipse brightness at that time. Therefore, a sector (S14) of the light curve approximately 28 days from TJD 1682 to 1710 is selected to analyse the basic model solution without spot. In order to remove the possible effect of brightness pulsations, the data from S14 were phased and then binned to make them suitable for binary modelling using the Wilson-Devinney (W-D) code (\citealp{1971ApJ...166..605W,1979ApJ...234.1054W,1990ApJ...356..613W,2007ApJ...661.1129V,2012AJ....144...73W}).

 In our W-D analysis, some parameters were set as free parameters, others were fixed. The limb darkening coefficients were taken from \cite{1993AJ....106.2096V} according to the wavelength of TESS and the temperature of the components.  The theoretical values of the bolometric albedos were taken from  \cite{2001MNRAS.327..989C} as to be 1.0 (primary star) and 0.5 (secondary star), corresponding to radiative and convective atmospheres, respectively. Gravity darkening exponents were selected as to be 1.0 for radiative atmospheres (primary component) and 0.32 for convective atmospheres (secondary component) from \cite{1967ZA.....65...89L}. The surface effective temperature of the primary star, 8103 K, was taken from our spectroscopic study. It was assumed that the components would rotate synchronously in a circular orbit, so the rotational parameters $(F_1,F_2)$ for the primary and secondary stars were set to 1.0 and the eccentricity $(e)$ was set to 0. As the radial velocity of the secondary star is lacking, the mass ratio $(q)$ of HZ Dra was selected with adjustable parameters during the analysis. The other adjustable parameters were the orbital inclination ($i$), the surface temperature of the secondary star ($T_2$) ,  the dimensionless potential of the primary ($\Omega_1$), the phase shift ($\phi$), the monochromatic luminosity of the primary component ($L_1$)) and the third light ($l_3$).
 
\begin{figure*}[ht!]
 	\plotone{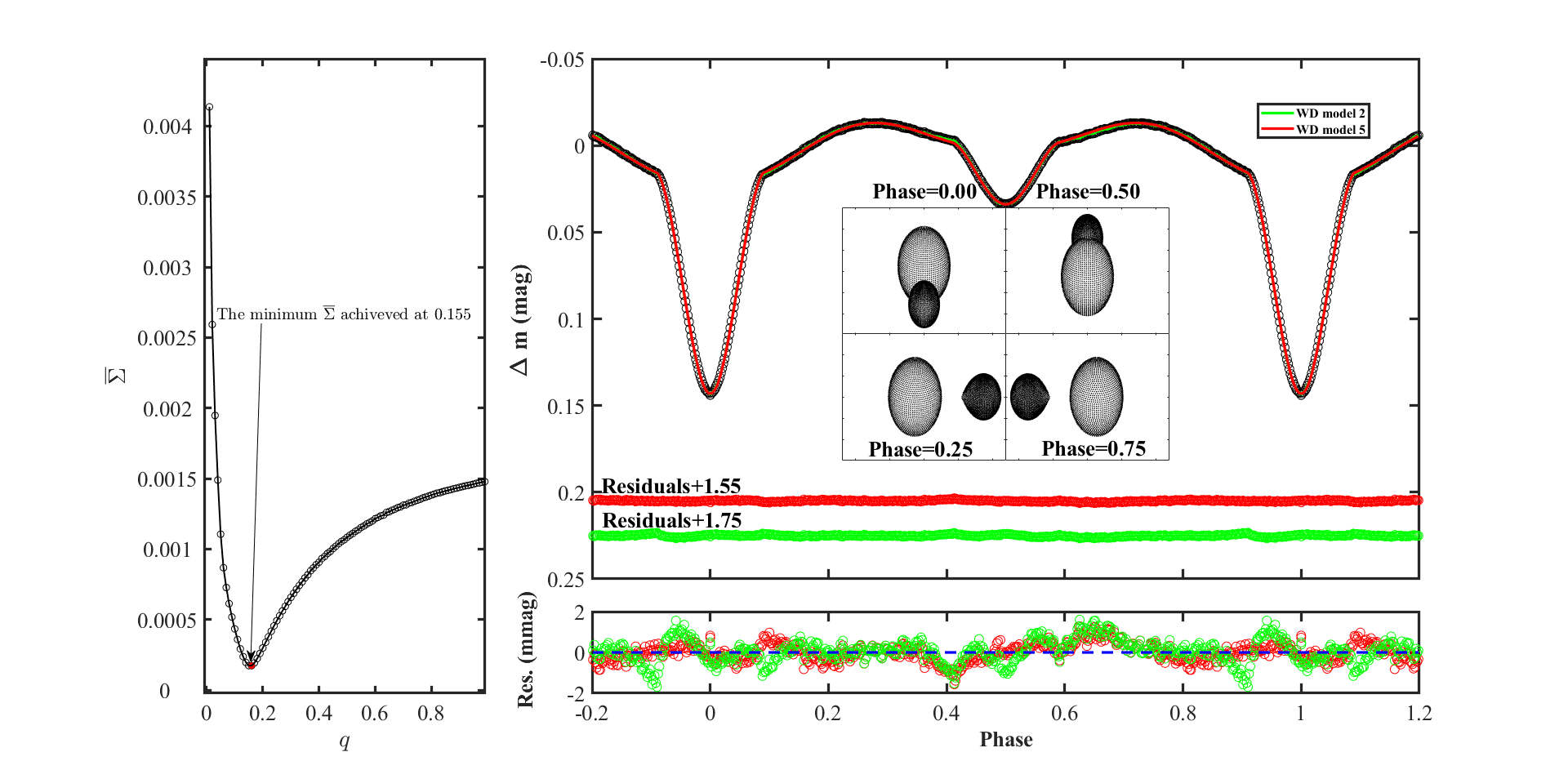}
 	\caption{The basic solutions for HZ Dra. Left panel: the relationship between $q$ and the mean residuals $\overline{\Sigma}$ for W-D Model 5, the red dot indicates the best $q$ with the minimum mean residual. Right upper panel: the average light curve (black circles), the theoretical fit curve (the red solid line for Model 5 and the green solid line for Model 2) and the Roche lobe geometry of HZ Dra at phases 0.00, 0.25, 0.50 and 0.75, respectively. Right lower panel: the enlarged view of residuals for average light curve (the red and green for Models 5 and 2, respectively). \label{fig:figure 4}}
 \end{figure*}

 For the photometric models, Model 2 was selected, which corresponds to a detached configuration in the W-D code. Dimensionless potential of the secondary ($\Omega_2$) is also selected as an adjustable parameter. After initial iterations, according to its dimensionless surface potential value, it was decided that the secondary component had filled its critical inner Roche lobe. Therefore, our photometric solution was carried out in Model 5, which means that a semi-detached configuration is assumed for HZ Dra. The iterations were continued until the corrections of the free parameters were smaller than their probable errors. In addition, a $q$-search method was carried out by checking the minimum $\overline{\Sigma}$ value. The resulting mean residuals $\overline{\Sigma}$ for each $q$ are plotted in left panel of Figure \ref{fig:figure 4}. The minimum value of $\overline{\Sigma}$ is achieved at $q$ = 0.155. Iterations were performed with the initial input parameters of $q$ = 0.155 until convergence. The final basic photometric solutions were derived and listed in the upper part of Table \ref{tab:table 2}. The observational light curve (the black circles), the theoretical light curve (the solid lines), and the theoretical geometrical structures are displayed in Figure \ref{fig:figure 4}.
 
\begin{table*}
	\begin{center}
		\footnotesize
		\caption{Photometric solutions of Model 5 using the W–D method and fundamental parameters for HZ Dra.\label{tab:table 2}}
		\begin{tabular}{llcc}\hline
			Parameters &&Values\\\hline
			&Photometric solutions&\\
			$g_1$&&1.00 (fixed)\\
			$g_2$&&0.32 (fixed)\\
			$A_1$&&1.00 (fixed)\\
			$A_2$ &&0.50 (fixed)\\
			$T_1$(K) &&8103 (fixed)\\
			$i$ ($^{\circ}$) &&68.92 $\pm$ 0.07\\
			$q$ ($M_2/M_1$)&&0.155 $\pm$ 0.001\\
			$T_2/T_1$&&0.582 $\pm$ 0.005\\
			$L_1/(L_1+L_2)_{TESS}$&&0.942 $\pm$ 0.025\\
			$L_3/(L_1+L_2+L_3)_{TESS}$ && 29 $\%$ $\pm$ 4$\%$\\
			$\Omega_1$&&2.85 $\pm$ 0.06\\
			$r_1(pole)$&&0.3697 $\pm$ 0.0049\\
			$r_2(pole)$&&0.2165 $\pm$ 0.0041\\
			$r_1(point)$&&0.3892 $\pm$ 0.0056\\
			$r_2(point)$&&0.3193 $\pm$ 0.0041\\
			$r_1(side)$&&0.3815 $\pm$ 0.0054\\
			$r_2(side)$&&0.2252 $\pm$ 0.0043\\
			$r_1(back)$&&0.3857 $\pm$ 0.0055\\
			$r_2(back)$&& 0.2571 $\pm$ 0.0044\\
			mean residuals &&0.00014\\\hline
			&Fundamental parameters&\\
			$M_1$ (M$_{\odot}$) &&1.64 $\pm$ 0.02\\
			$M_2$ (M$_{\odot}$) &&0.255 $\pm$ 0.004\\
			$R_1$ (R$_{\odot}$) &&1.67 $\pm$ 0.01\\
			$R_2$ (R$_{\odot}$) &&1.11 $\pm$ 0.01\\
			$L_1$ (L$_{\odot}$) &&10.84 $\pm$ 1.35\\
			$L_2$ (L$_{\odot}$) &&0.55 $\pm$ 0.07\\ 
			$M_{\rm bol1}$ &&2.15 $\pm$ 0.13\\
			$M_{\rm bol2}$ &&5.38 $\pm$ 0.14\\
			$a$ (R$_{\odot}$) &&4.39 $\pm$ 0.02\\
			\hline
		\end{tabular}
	\end{center}
\end{table*}

Combining the radial velocity measurements for the primary star of \cite{2012MNRAS.422.1250L} and our photometric solutions, the fundamental stellar parameters, such as mass ($M$), radius ($R$), luminosity ( $L$ ), bolometric magnitude ( $M_{\rm bol}$ ) and separation between the components ( $a$ ), for both binary components were calculated using the mass function \citep{2001icbs.book.....H}, the well-known Kepler and Pogson equations. These parameters are listed in lower panel of Table \ref{tab:table 2}.

The O$^{'}$Connell effect in HZ Dra is visible, which probably caused by one or more star spots. As shown in the lower panel of Figure \ref{fig:figure 3}, because the changing O$^{'}$Connell effect from TJD 2600-3000 is stronger than that from TJD 1790-1810 and has a long duration. Four TESS sections of the light curves from TJD 2580-2605, TJD 2665-2693, TJD 2795-2825 and TJD 2940-2965, were therefore selected to analyze the variations of O$^{'}$Connell effect. We found that if a hot spot is added to the primary star and another cool spot is added to the secondary star, all four light curves can be fitted well.
\begin{figure*}[ht!]
	\centering
	\begin{minipage}{1\linewidth} 
		\setlength{\abovecaptionskip}{-4pt}
		\subfigure{
			\label{fig:1}
			\includegraphics[width=0.50\linewidth,height=3.0in]{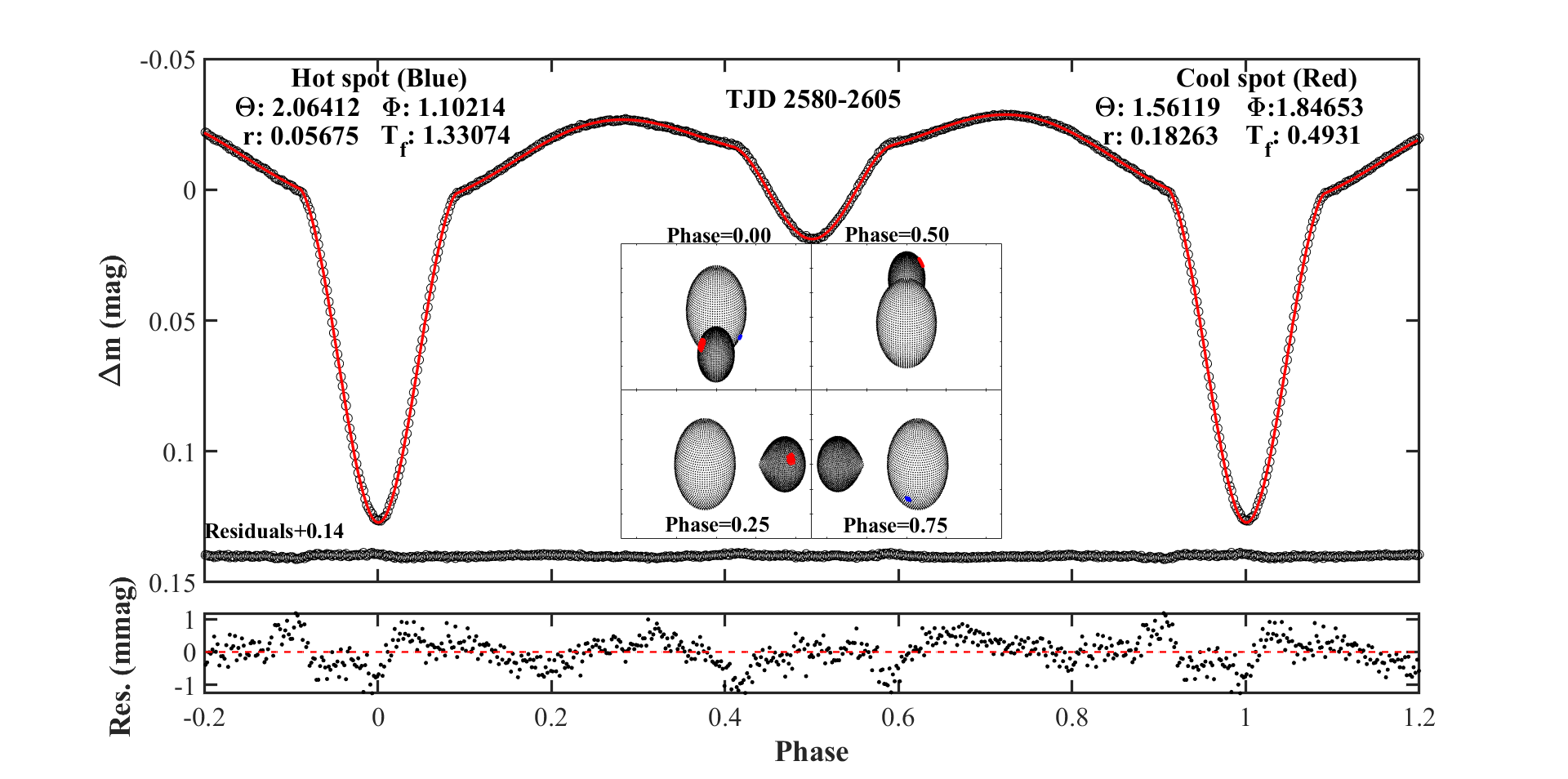} 
		}
		\subfigure{
			\label{fig:2}
			\includegraphics[width=0.50\linewidth,height=3.0in]{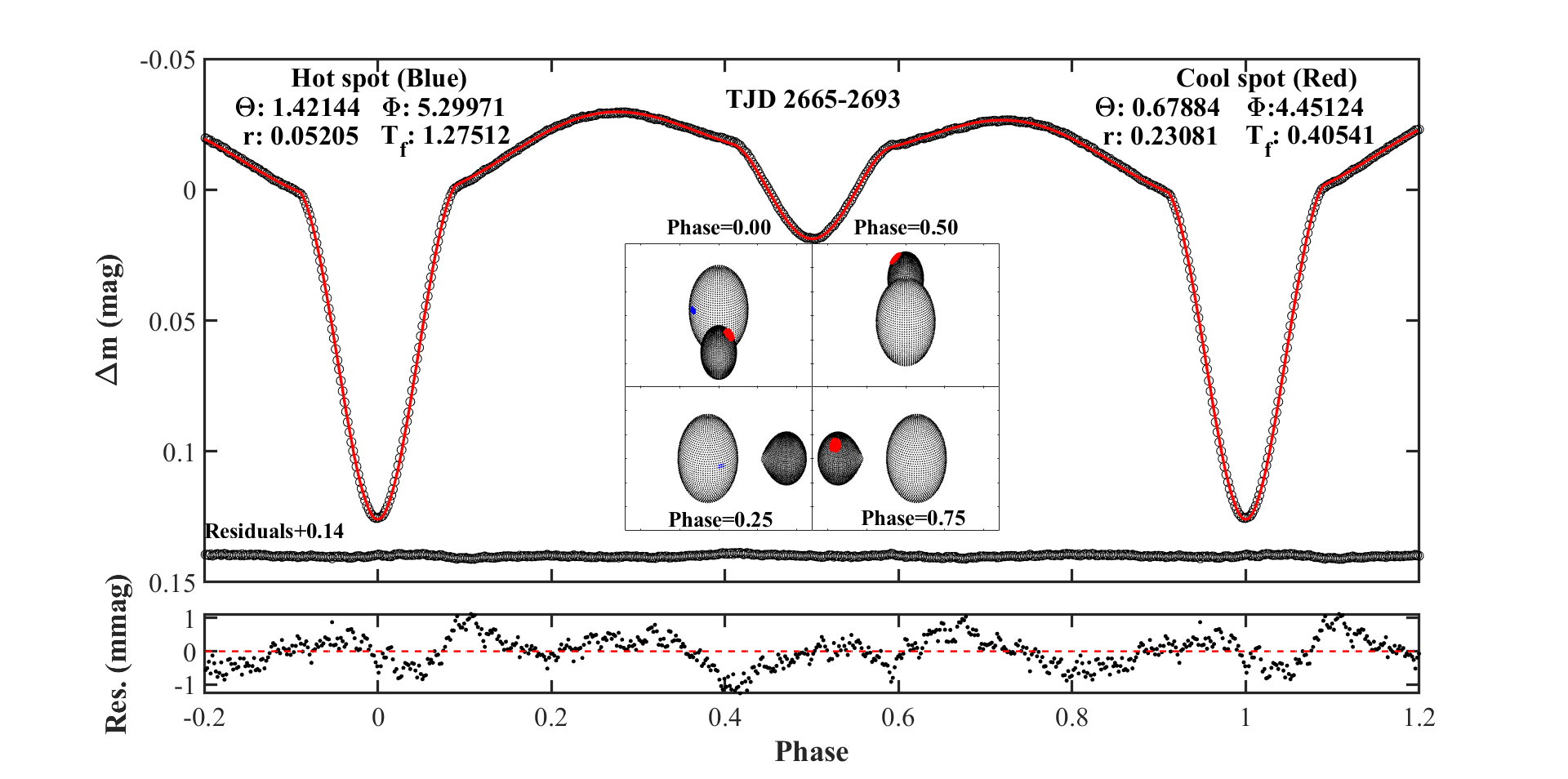}
		}
		\subfigure{
			\label{fig:1}
			\includegraphics[width=0.50\linewidth,height=3.0in]{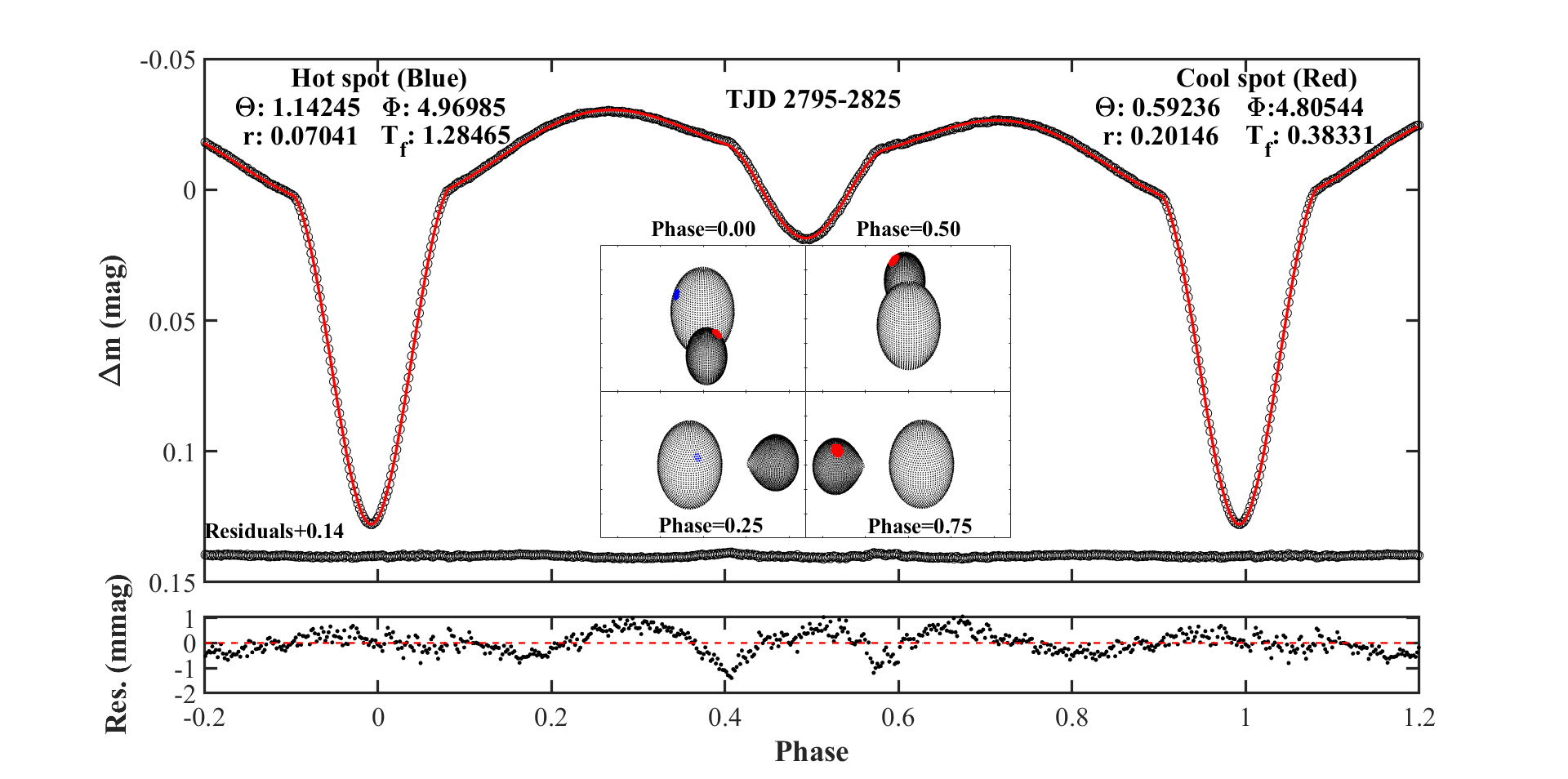} 
		}
		\subfigure{
			\label{fig:2}
			\includegraphics[width=0.50\linewidth,height=3.0in]{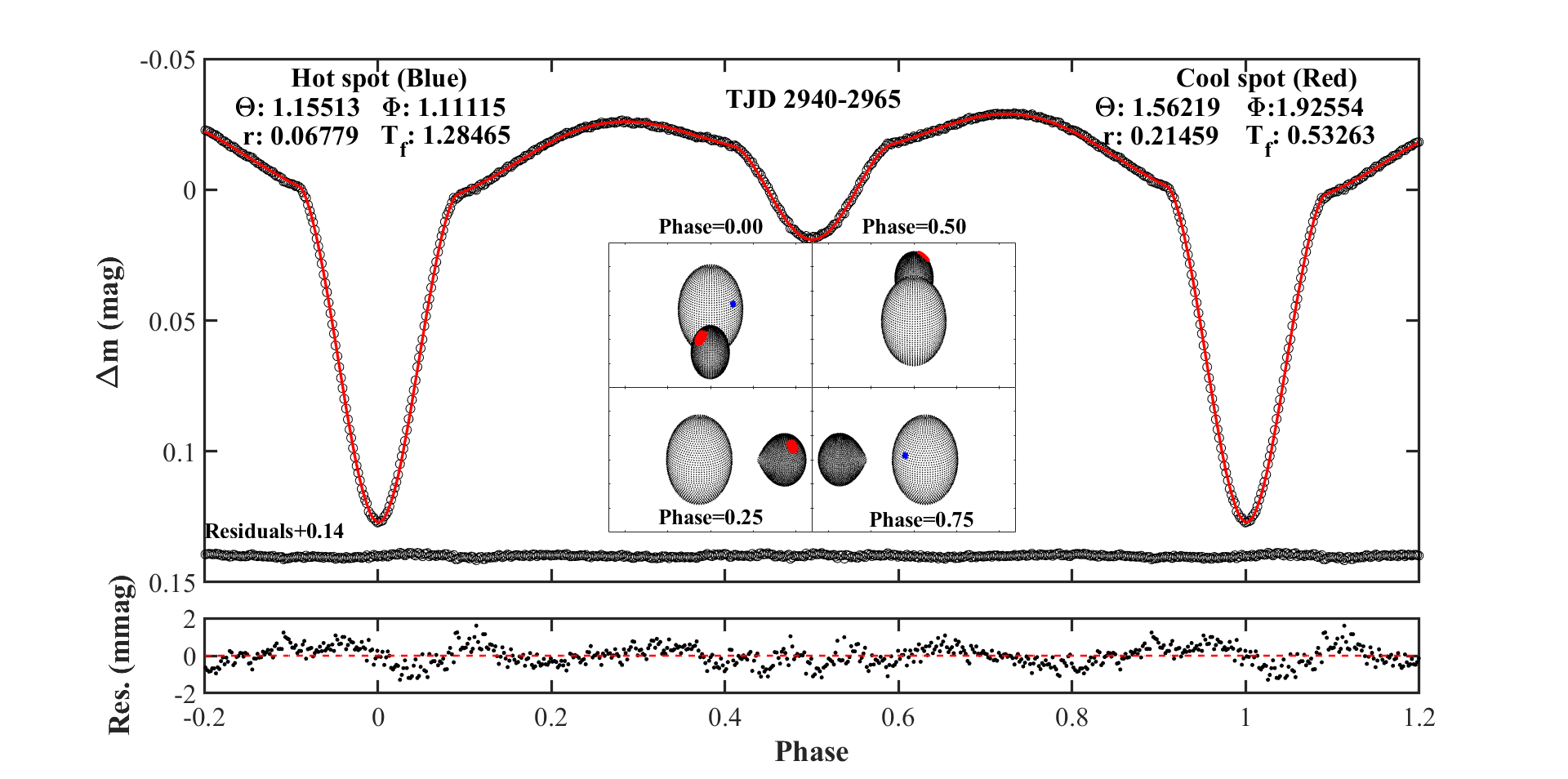}
		}
	\end{minipage}
	\caption{The graphs of spot solutions obtained using the W-D program. Top panels: the average light curve (black circles), the calculated fitting light curves (red solid line), and the Roche lobe geometry. Spot parameters are also marked in these panels: latitude ($\theta$, radian), longitude ($\phi$, radian), angular radian ($r$, radian), temperature factor ($T_f$, the ratio of spot temperature to stellar temperature). Bottom panels: the enlarged views of the residuals. \label{fig:figure 5}}
	\vspace{0.0in}  
\end{figure*}

 As shown in Figure \ref{fig:figure 5}, the observed light curves (the black circles), the fitted light curves (the red solid lines) and the geometric structures are plotted. The parameters of spots are also shown in the figure. We found that the positions of the cool and hot spots are evolutionary and roughly symmetric with the inner Lagrangian point L1 between two components. Therefore the change of the O$^{'}$Connell effect can be caused by the evolutionary spots, the same phenomenon was also observed in KIC 06852488 \citep{2021AJ....161...46S}.
 
 The volume filling factors ($V_{\rm star}/V_{\rm L}$) of the primary and the secondary are 34.11 ($\pm$0.15)$\%$ and 98.45 ($\pm$0.67)$\%$, respectively. The hot spot thus may be caused by the mass transfer from the secondary component to the primary star. In addition, the cool spot may be caused by magnetic activity on the surface of the secondary component, because it is a late-type star. 

\section{Orbital period analysis} \label{sec:OC}
Orbital period analysis has been neglected for HZ Dra. Actually, it is very useful to analyze the period variation for oEA, as it can provide information on the mass transfer, magnetic activity, additional bodies around the central binaries and their possible effect on the pulsation behavior of the massive component. The orbital period variation of HZ Dra was presented in our work.

Using the TESS data, 1011 primary eclipsing times and 1003 secondary eclipsing times were calculated using the method of \cite{1956BAN....12..327K}. HZ Dra was also observed with the 60 cm Cassegrain reflecting telescopes at the Yunnan Observatories (YNOs) of CAS on 05 March 2021 and 21 May 2023 using a standard Johnson–Cousins multi-color BVR$_c$I$_c$ filter system. The telescope is equipped with Andor DW436 2048 $\times$ 2048 CCD. The CCD images were reduced using the DAOPHOT package of the IRAF software in a standard process that includes bias and flat. We thus obtained 3 primary eclipsing times from the 60 cm telescope light curves. A sample of eclipsing times are listed in Table \ref{tab:table 3}.
\begin{table*}
	\begin{center}
		\footnotesize
		\caption{Eclipsing times of HZ Dra. This table is available in full in machine-readable form. \label{tab:table 3}}
		\begin{tabular}{lllllc}\hline
			Eclipsing times (P) &Errors&Eclipsing times (S) &Errors&Source\\
			(HJD)&&(HJD)\\\hline
			2458683.40420 & 0.00023 & 2458683.79020 & 0.00086& TESS\\
			2458684.17720 & 0.00027 & 2458684.56420 & 0.00057& TESS\\
			2458684.95020 & 0.00020 & 2458685.33820 & 0.00055& TESS\\
			2458685.72320 & 0.00019 & 2458686.11020 & 0.00052& TESS\\
			2458686.49620 & 0.00022 & 2458686.88320 & 0.00037& TESS\\
			2458687.26920 & 0.00020 & 2458687.65520 & 0.00064& TESS\\
			2458688.04220 & 0.00021 & 2458688.42920 & 0.00029& TESS\\
			2458688.81420 & 0.00019 & 2458689.20220 & 0.00034& TESS\\
			2458689.58720 & 0.00026 & 2458689.97420 & 0.00037& TESS\\
			2458690.36020 & 0.00020 & 2458690.74720 & 0.00049& TESS\\
			
			... & ... & ... & ...&...\\ 
			2459289.391 & 0.002 & ... & ...& YNOs\\
			2459330.359 & 0.002 & ... &...& YNOs\\
			2460086.286 & 0.001 & ... & ...& YNOs\\
			\hline
		\end{tabular}
	\end{center}
\end{table*}

\begin{figure*}[ht!]
	\plotone{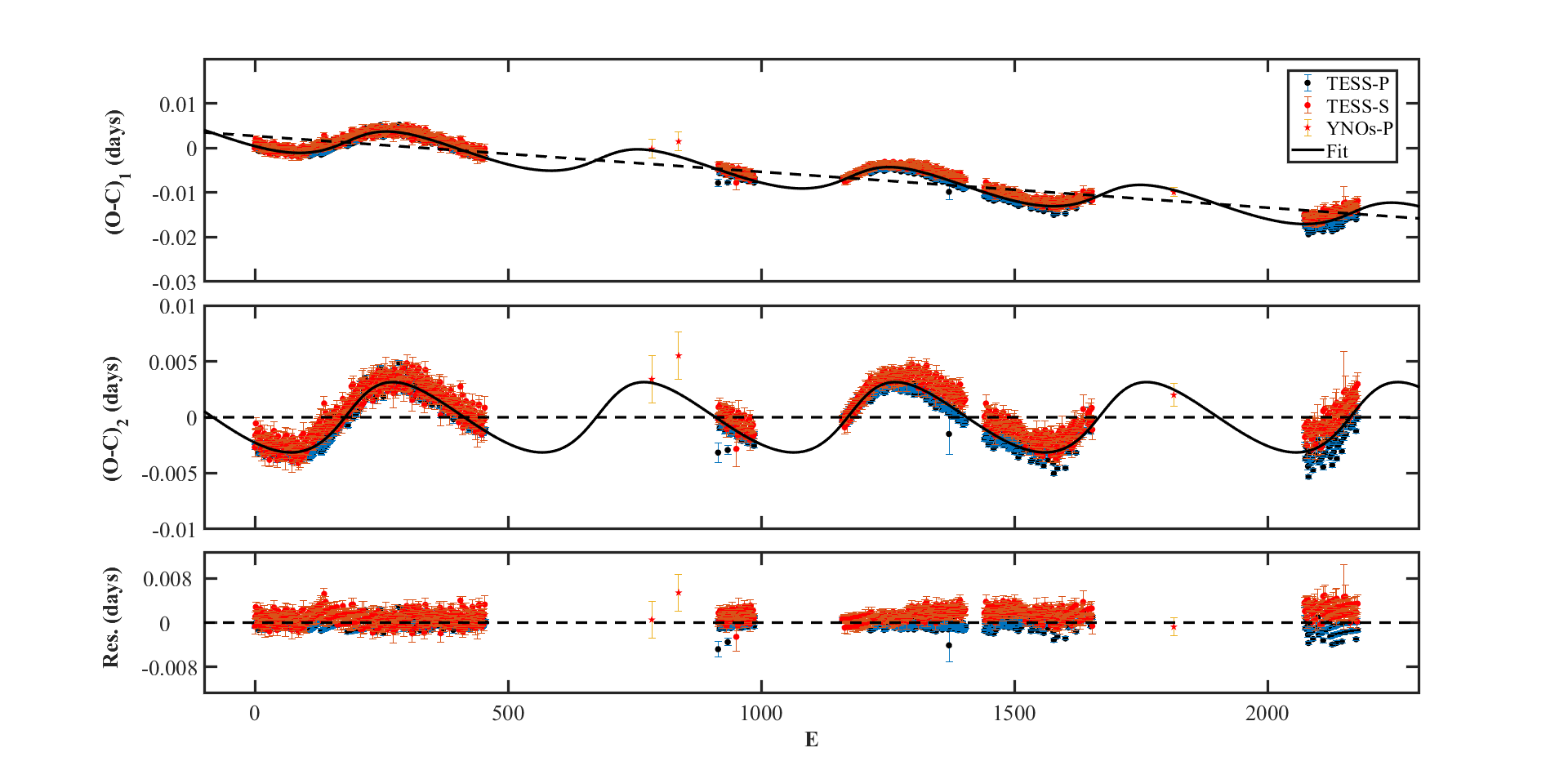}
	\caption{The $O-C$ diagrams of HZ Dra. Top panel: the $(O-C)_{1}$  values calculated by using Equation (\ref{equation(1)}). The black dashed and solid lines indicate the model of a slash line plus LTTE (Light travel time effect), the slash line model indicates that the ephemeris needs to be revised. The red and black dots refer to data from TESS, and the red pentagrams to data from the 60 cm telescope of the Yunnan Observatories of CAS. Middle panel: The $(O-C)_{2}$ data of the mainly cyclical variation after subtracting the linear change from the $(O-C)_{1}$ values. Bottom panel: the variations of the residual data of the middle panel.\label{fig:figure 6}}
\end{figure*}

The following linear ephemeris, where the primary eclipsing time of TESS and the orbital period given by \cite{2002SASS...21....9C}, is used to calculate the $O-C$ values and epoch numbers ($E$): 
\begin{eqnarray}\label{equation(1)}
	MinI (\rm {HJD}) =2458683.40420 + 0^{\textrm{d}}.7729340\times{E}.
\end{eqnarray}
As shown in the top panel of Figure \ref{fig:figure 6}, the distribution of $(O-C)_{1}$ data shows a linear plus periodic variation. The cyclic variation can be explained in two ways:  (1) light travel-time effect (LTTE) and (2) magnetic activity (Applegate mechanism).

\subsection{Light travel-time effect}
The existence of third body around the binary system, causing a periodic variation in the $O-C$ diagrams of the eclipsing binaries. This variation can be interpreted by LTTE. Many possible potentional companions are found around oEA systems using $O-C$ analysis (c.g., \citealp{2013AJ....145...87S}; \citealp{2016RAA....16...94L}; \citealp{2021MNRAS.505.6166S}; \citealp{2022MNRAS.510.1413K}).

We fit the $(O-C)_1$ variations using the following equations \citep{1952ApJ...116..211I}, a model of a slash line plus LTTE:
\begin{eqnarray} \label{equation(2)}
	O-C (\rm d)&=&\Delta T_0+\Delta P_0\times E+A[(1-e_{3}^{2})\frac{\sin(\nu+\omega)}{1+e_{3}\cos\nu}+e\sin\omega]\nonumber\\
	&=&\Delta T_0+\Delta P_0\times E+A[\sqrt{1-e_{3}^{2}}\sin E^{*}\cos \omega+\cos E^{*}\sin \omega],
\end{eqnarray}
and the Kepler's equation:
\begin{eqnarray} \label{equation(3)}
	M&=&E^{*} - e_{3} \sin E^{*}=\frac{2\pi}{P_{3}} (t - T_{3}).
\end{eqnarray}
In the above equations, the $\Delta T_0$ and $\Delta P_0$ are the revised epoch and period, the $A$ is the semi-amplitude of $O-C$ variation. Meanwhile, the $e_{3}$, $\nu$ and $\omega$ are the eccentricity, true anomaly and longitude of the periastron from the ascending node, the $M$, $T_{3}$ and $t$ are the mean anomaly, time of periastron passage and observed  eclipsing times, the $P_{3}$ and $E^*$ are the period and the eccentric anomaly. 
\begin{figure*}[ht!]
	\plotone{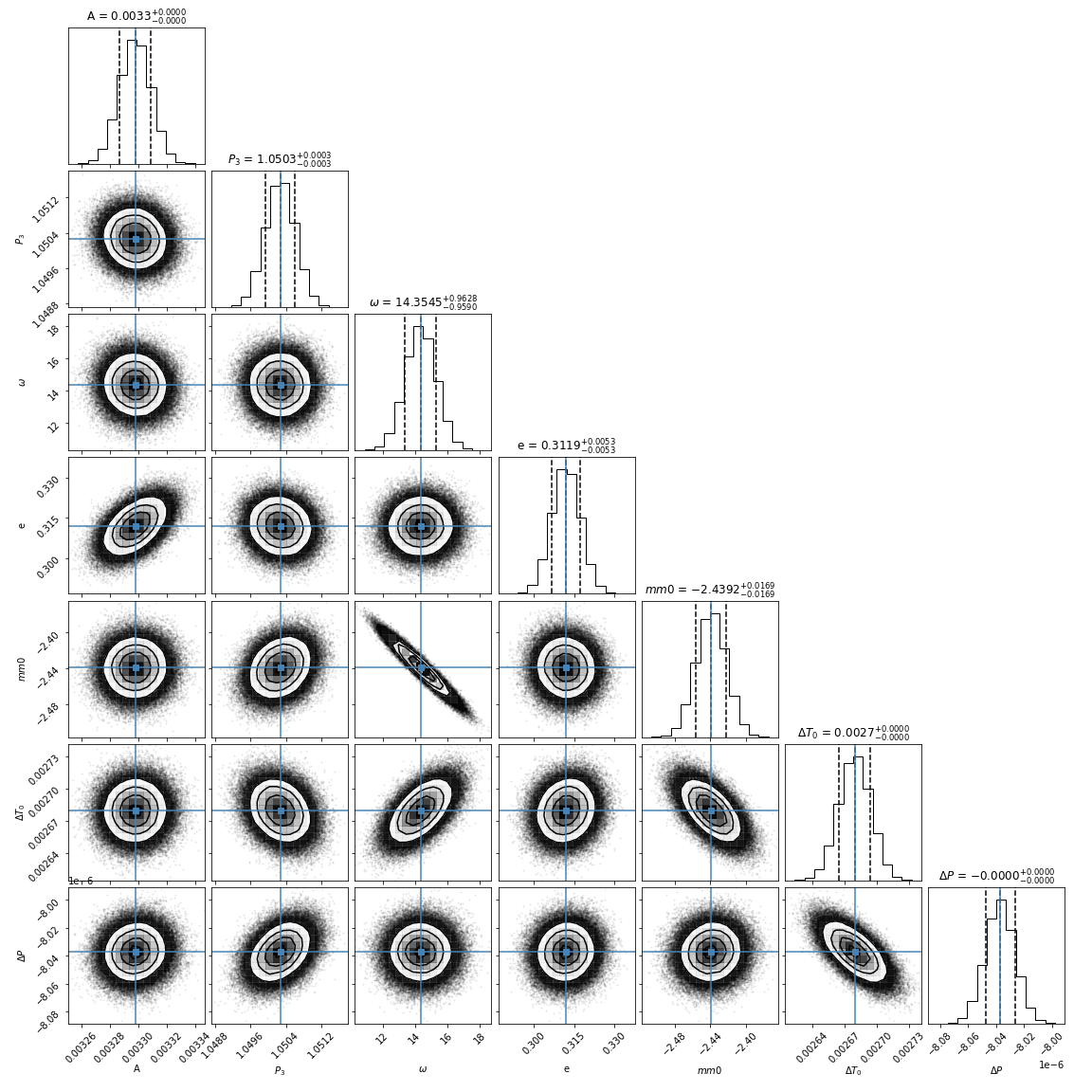}
	\caption{Corner plot of the MCMC fitting code for the model of a slash line plus LTTE. Blue vertical lines indicate the median values of the histograms presented for each parameter. Posterior distributions at the 16 per cent and 84 per cent quantiles are shown by black vertical dashed lines. The units for $A$, $P_{3}$, $\omega$, $\Delta T_0$, $\Delta P_0$ and $d$ are days, years, degree, HJD and days, respectively. In the plot, $mm0=\frac {2\pi(2458683.4042-T_{3})}{365.25\times P_{3}}$.\label{fig:figure 7}}
\end{figure*}

The Markov Chain Monte Carlo (MCMC) method with the $emcee$ v3.1.2 Python package \citep{2013PASP..125..306F} was used to fit the (O-C)$_1$ based on the above model. Because the method has excellent performance and good programming flexibility. In the MCMC run, a physically reasonable range limits for each parameter of the model are set as follows: 0 $<A<$ 0.1 (days), 0 $< P_{3} <$ 3 (years), -360$^{\circ}$ $<$ $\omega$ $<$ 360$^{\circ}$, 0 $< e <$ 1, -10 $< mm0 <$ 10,  -0.1 $< \Delta T_0 <$ 0.1 (HJD), -0.1 $< \Delta P_0 <$ 0.1 (days), where the parameter $mm0$ is relate to $T_{3}$. In the first round of the fitting process, we set all parameters as an uniform distribution of the above range. Then we use 50 walkers and 5000 steps for a single chain. After this round, we could obtain proper parameters. Finally, following the similar method of \cite{2023ApJS..266...28L}, we derived the uncertainties of the parameters by using 100 walkers and 10,000 steps in a single chain. Also, their priors are set to a normal distribution, using the values from the previous step as the expectation and its 0.1 times as the variance. The corresponding corner plot of the MCMC fitting program for the model is shown in Figure \ref{fig:figure 7}, and there are clearly correlations between many parameters, for example, $\omega-e$, $mm0-\Delta T_0$ and $\Delta T_0-\Delta P$. The fitting parameters of this model are listed in Table \ref{tab:table 4}.
\begin{table*}
	\begin{center}
		\footnotesize
		\caption{The fitting parameters of linear plus LTTE obtained from $O-C$ analysis by using MCMC method.\label{tab:table 4}}
		\begin{tabular}{llcc}\hline
			Parameters&Value\\\hline
			$A$ (days)&0.0033 $\pm$ 0.0001  \\
			$\omega(^o)$&14.35 $\pm$ 0.95 \\
			$e$&0.311 $\pm$ 0.005 \\
			$P_{3}$ (years)&1.0503 $\pm$ 0.0003 \\
			$T_{3}$ (HJD)&2458832.3213 $\pm$ 1.0338 \\
			$\Delta T_{0}$ (HJD)&0.00267 $\pm$ 0.00146\\
			$\Delta P_{0}$ (days)&0.0000009  $\pm$0.0000001 \\
			mean residuals&0.0053\\
			\hline
		\end{tabular}
	\end{center}
\end{table*}

From the middle panel of Figure \ref{fig:figure 6}, we can see clearly 
LTTE caused by a possible third body. The parameters of the third body are calculated by using the following equation
\begin{eqnarray}\label{equation(4)}
	f(m) =\frac{4\pi^2}{GP^2_3}\times(a^{\prime}_{12}\sin{i^{\prime}})^3=\frac{(M_3\sin{i^{\prime}})^3}{(M_1+M_2+M_3)^2},
\end{eqnarray}
where $a^{\prime}_{12}$$\sin{i^{\prime}}$ = $A\times c$ (c is the speed of light ) is the projected semi-major axis of the hypothetical triple orbit, which was calculated to be 0.92 $\pm$ 0.03 au with an inclination of 90$^{\circ}$. The orbital period ($P_3$) of the third body around the mass center of the triple system was determined to be 1.0503 $\pm$ 0.0003 yr. The mass function of the third component can thus be calculated to be 0.17 $\pm$ 0.01 M$_{\odot}$. So the minima mass of the third body was calculated to be 1.16 $\pm$ 0.01 M$_{\odot}$. This value of the tertiary component is much larger than the mass of the secondary component, it was not detected in previous studies.

\subsection{Applegate mechanism}
Most of the cooler components of classical Algols are late-type stars with deep convective layers.  If one of the components in a binary system has magnetic activity, the orbital angular momentum of the system can be changed. \cite{1992ApJ...385..621A} suggested that periodic variation in $O-C$ curves are due to the magnetic activity of a late-type cooler component in eclipsing binaries. 

\begin{figure*}[ht!]
	\plotone{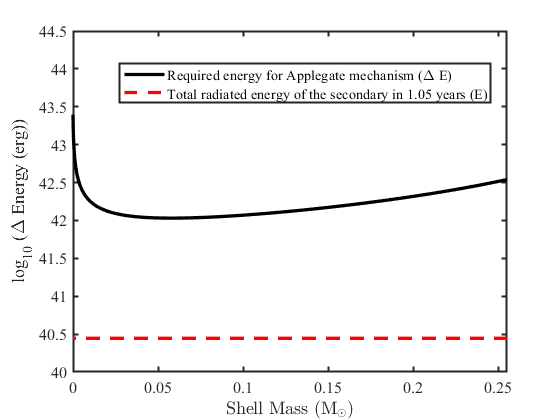}
	\caption{The energy required to produce the periodic variation in the $O-C$ diagram using the Applegate mechanism model.\label{fig:figure 8}}
\end{figure*}

In order to estimate the possible affect of Applegate mechanism on the period variation of HZ Dra, we used the method reported by \cite{2006MNRAS.365..287B} to calculate the energies required to cause this periodic variation. As shown in Figure \ref{fig:figure 8}, for the secondary star of HZ Dra, the energy required for the Applegate mechanism is greater than the total radiative energy in an entire period. This means that the secondary is unable to trigger Applegate mechanism. Additionally, the $O-C$ diagram shows a strictly periodic variation. We therefore prefer to the LTTE explanation.

\section{Frequency analysis} \label{sec:Fourier}
By subtracting the tenth-order Fourier series fit of the orbital frequency (i.e. the multi-frequency harmonic mode of the orbital frequency; see e.g. \cite{2021MNRAS.501L..65S}) from the TESS light curves of S14-26, we obtained light residuals.
\begin{figure*}[ht!]
	\plotone{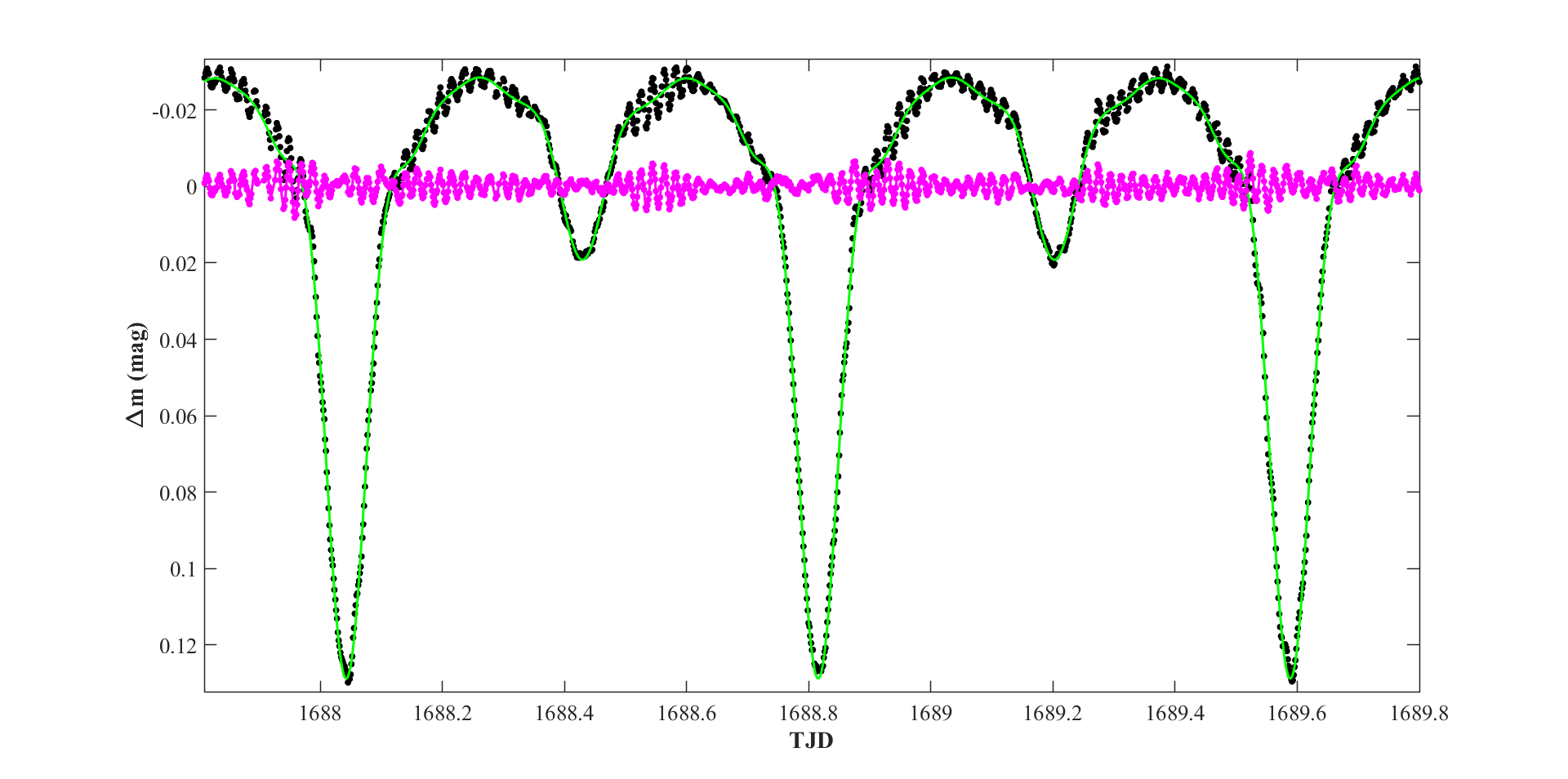}
	\caption{One sector of the TESS light curve fitted by the tenth-order Fourier series of the orbital frequency. The black dots refer to the TESS light curve. The green solid line represents the fitting light curve. The pink dots refer to the fitting residuals.\label{fig:figure 9}}
\end{figure*}
Figure \ref{fig:figure 9} presents one sector of light curve and its light residuals in plot of magnitude versus TJD. To study the pulsation features in detail, we performed a multiple frequency analysis of the light residuals with the Period04 software \citep{2005CoAst.146...53L}. Since no pulsation signals are detected in the high frequency region (f $>$ 120 d$^{-1}$), further frequency extraction was performed in the frequency range 0-120 d$^{-1}$. 
\startlongtable
\begin{deluxetable*}{lcccclcccc}
	\tabletypesize{\scriptsize}
	\label{tab:table 5}
	\tablecaption{The pulsating frequencies of HZ Dra. The numbers in parentheses are the errors on the last bit of the data and the capital letters A-F in brackets represent 6 multiplets. The $F_o$=1.29375433 d$^{-1}$  is orbital frequency.}
	\tablehead{
		\colhead{ID}& \colhead{Freq.(d$^{-1}$)} & \colhead{Ampl.(mag)} & \colhead{Remark} &\colhead{S/N}  &\colhead{ID}& \colhead{Freq.(d$^{-1}$)} & \colhead{Ampl.(mag)} &\colhead{Remark}  & \colhead{S/N}
	} 
	\startdata
	$F_1$  & 52.069971 (42) & 0.0028866 (57) & $(A)$ &234.11 & $F_{50}$	&52.6924 (15)&0.0000804 (57)&$254F_o-5F_{13}$&7.8\\
	$F_2$  & 57.31032 (11) &  0.0010639 (57) &$(B)$ &93.41&              $F_{51}$	&20.7020 (15)&0.0000792 (57)&16F$_o$&13.43\\ 
	$F_3$  & 48.98764 (12) &  0.0009831 (57)  & $(C)$ &95.68 &          $F_{52}$	&10.3665 (15)&0.0000780 (57)&$8F_{20}-309F_o$&8.51\\
	$F_4$  & 54.16202 (13) & 0.0009230 (57) &  $F3+4F_o(C)$& 76.28&       $F_{53}$	&9.0462 (13)&0.0000887 (57)&$31F_o-4F_{32}$&11.77\\
	$F_5$  & 51.21916 (12) & 0.0009384 (57)  & $2F2-49F_o$& 103.23&       $F_{54}$	&14.2330 (16)&0.0000756 (57)&$11F_o$&9.75\\ 
	$F_6$  & 45.14224 (15) & 0.0007916 (57) &  $(D)$    & 92.91&          $F_{55}$	&1.2290 (16)&0.0000730 (57)&$121F_o-20F_{32}$&5.7\\ 
	$F_7$  & 49.48233 (17) & 0.0006932 (57) & $F_1-2F_o(A)$& 65.01&        $F_{56}$	&0.1792 (15)	&0.0000775 (57)&$38F_o-F_3$&6.9\\
	$F_8$  & 54.65734 (19) & 0.0006323 (57) & $F_1+2F_o(A)$& 45.12&       $F_{57}$	&47.5107 (14)	&0.0000843 (57)&$2F_3-39F_o$&8.80\\                  
	$F_9$  & 54.87759 (23) & 0.0005186 (57)  &       & 37.73&             $F_{58}$	&49.6165 (16)	&0.0000751 (57)&$81F_o-F_{13}$&7.76\\                   
	$F_{10}$ & 47.73033 (32) & 0.0003702 (57)& $F6+2F_o(D)$&34.70&        $F_{59}$	&55.7839 (17)	&0.0000702 (57)&$78F_o-F_6$&6.32\\
	$F_{11}$ & 1.29877 (13) & 0.0009158 (57)  &$315F_o-9F_6$& 71.85&        $F_{60}$	&45.0753 (16)	&0.0000746 (57)&$Sidelobe$&8.49\\                   
	$F_{12}$ & 48.63228 (34) & 0.0003515 (57) & $2F2-51F_o$& 30.86&         $F_{61}$	&1.30477 (19)&0.0006263 (57)&$4F_{32}-23F_o$&48.79\\                
	$F_{13}$ & 55.19183 (41) & 0.0002956 (57)  & & 23.29&                   $F_{62}$&50.2809 (19)&0.0000628 (57)&$F_3+F_o(C)$&	6.89\\                  
	$F_{14}$ & 51.57400 (45) & 0.0002676 (57)  & $F_3+2F_o(C)$& 25.67&       $F_{63}$&56.0183 (19)&0.0000614 (57)&$F_2-F_o(B)$&	5.80\\                  
	$F_{15}$ & 52.60679 (41) & 0.0002894 (57) &$F_{13}-2F_o$& 25.94&           $F_{64}$	&60.2295 (19)&	0.0000612 (57)	&$7F_{29}-234F_o$&	5.27\\            
	$F_{16}$ & 53.86603 (33) & 0.0003668 (57)&$9F_2-357F_o$&32.95&           $F_{65}$	&15.5182 (20)&	0.0000591 (57)&$9F_6-302F_o$&	8.59\\ 
	$F_{17}$ & 53.80641 (36) & 0.0003341 (57)&$2F_2-47F_o$& 29.76&           	$F_{66}$	&58.6045 (20)&	0.0000604 (57)&$F_2+F_o(B)$&	6.19\\                           
	$F_{18}$ & 42.55443 (49) & 0.0002463 (57)&$F6-2F_o(D)$&22&          $F_{67}$	&57.7197 (20)	&0.0000585 (57)	&$245F_o-5F_{29}$&	4.55\\            
	$F_{19}$ & 2.59360 (34)  & 0.0003559 (57)&$316F_o-9F_6$&30.81&          $F_{68}$	&1.4829 (20)	&0.0000616 (57)	&$60F_{32}-359F_o$&4.62\\            
	$F_{20}$ & 51.27316 (46) & 0.0002633 (57)&$Sidelobe$ & 27.73&           $F_{69}$	&50.7753 (21)	&0.0000576 (57)	&$F_1-F_o(A)$&	6.01\\
	$F_{21}$ & 45.73240 (58) & 0.0002062 (57)&  $(E)$& 22.30&             $F_{70}$	&55.16847 (17)	&0.0000680 (57)&$Side lobe$&5.42\\             
	$F_{22}$ & 46.00899 (69) & 0.0001753 (57) & $187F_o-4F_3$& 17.30&     $F_{71}$&	9.0656 (15)&	0.0000791 (57)&$321F_o-9F_{60}$&10.45\\             
	$F_{23}$ & 6.46171 (68) & 0.0001767 (57) & $9F_6-309F_o$& 18.31&      $F_{72}$&	52.1034 (20)&	0.0000582 (57)&	$Side lobe$	&4.58\\             
	$F_{24}$ & 52.51561 (70) & 0.0001724 (57)& $2F_2-48F_o$& 15.43&      $F_{73}$&	52.8672 (16)&	0.0000733 (57)&	$F_3+3Fo(C)$&	7.32\\            
	$F_{25}$ & 53.49714 (76) & 0.0001589 (57) & $84F_o-F_{13}$& 16.02&      $F_{74}$&	53.1995 (18)	&0.0000668 (57)	&$338F_9-7F_o$&	7.26\\            
	$F_{26}$ & 51.39426 (75) & 0.0001608 (57) &$2F_3-26F_o$& 17.39&      $F_{75}$&	3.86045 (13)	&0.0000928 (57)&$	39F_o-6F_{32}$&	8.33\\              
	$F_{27}$ & 48.37459 (79) & 0.0001528 (57)&$7F_{20}-240F_o$& 13.91&     $F_{76}$&	0.2641 (19)	&0.0000625 (57)&$Side lob$&5.57\\                
	$F_{28}$  & 0.12708 (85) & 0.0001411 (57) &$9F_{21}-3180F_o$& 13.90&    $F_{77}$ &	51.0487 (23)	&0.0000525 (57)&$6F_1-202F_o$&5.67\\            
	$F_{29}$ & 51.85771 (88) & 0.0001365 (57) &$(F)$ &13.08&               $F_{78}$&	50.0945 (21)	&0.0000560 (57)&$	268F_o-16F_{76}$&	5.53\\          
	$F_{30}$ & 54.44501 (89) &0.0001352 (57)&$2F_o+F_{29}(F)$& 10.70&        $F_{79}$&	50.0023 (20)	&0.0000588 (57)	&$66F_o-134F_{76}$&	5.95\\
	$F_{31}$ & 52.8272 (10) & 0.0001352 (57)& $Side lobe$& 11.26 &      $F_{80}$&	53.9027 (23)	&0.0000511 (57)&$78F_o-178F_{76}$&	4.66\\
	$F_{32}$ & 7.766591 (92) & 0.0013176 (57)&  & 165.53 &             $F_{81}$&	42.4711 (24)&	0.0000504 (57)&	$2F_9-52F_o$&4.57 \\ 
	$F_{33}$ & 3.87605 (84) & 0.0001440 (57)&$9F_6-311F_o$ & 13.04&     $F_{82}$&	56.6506 (25)&	0.0000470 (57)&	$288F_o-7F_6$&	4.38  \\          
	$F_{34}$ & 5.1598 (11) & 0.0001030 (57) &$321F_o-8F_{20}$ & 11.39&     $F_{83}$&	0.3594 (24)& 	0.0000497 (57)&	$76F_o-2F_3$&4.69\\              
	$F_{35}$ & 49.7521 (11) & 0.0001033 (57)&$319F_o-7F_{29}$ & 11.11&     $F_{84}$&	58.9848 (24)&	0.0000486 (57)&	$2F_2-43F_o$&4.68  \\            
	$F_{36}$ & 49.0330 (11) & 0.0001042 (57) &$4F_o+F_{29}(F)$ & 9.70&     $F_{85}$&	5.20244 (22)&	0.0000528 (57)&	$7F_3-261F_o$&6.23 \\           
	$F_{37}$ & 2.58038 (53) & 0.0002275 (57)&$9F_6-312F_o$ & 19.54 &     $F_{86}$&	16.8153 (26)	&0.0000455 (57)&	$9F_6-301F_o$&6.26 \\          
	$F_{38}$ & 53.1467 (14) & 0.0000843 (57)&$31F_{32}-145F_o$ & 9.70 &     $F_{87}$&48.7105 (24)	&0.0000485 (57)	&$6F_o+155F_{76}$&4.16 \\            
	$F_{39}$ &47.1691 (10) & 0.0001103 (57)&$6F1-205F_o$ & 011.31  &    $F_{88}$&46.4388 (26)	&0.0000460 (57)	&$F_o+F_6(D)$    &4.13 \\          
	$F_{40}$ & 38.8191 (11) & 0.0001104 (57)&$30F_o$ & 12.41  &    $F_{89}$	&62.2068 (27)	&0.0000439 (57)	&$8F_o+F_{29} (F)$	   &6.68 \\  
	$F_{41}$ & 49.9317 (15) & 0.0000805 (57)&$81F_o-F_9$ & 8.26&        	$F_{90}$	&56.74839 (25)	&0.0000474 (57)	&$6F_o+F3(C)$&4.56\\           
	$F_{42}$ & 48.8044 (12) & 0.0000940 (57)&$2F_3-38F_o$ & 8.65&        	$F_{91}$	&57.3771 (25)	&0.0000471 (57)	&$87F_o-F_{13}$	&4.20\\              
	$F_{43}$	&11.6490 (13)&0.0000895 (57)&$F_{31}-32F_o$&12.02 &        $F_{92}$	&47.6543 (25)	&0.0000469 (57)	&$214F_o-4F_2$&4.65 \\              
	$F_{44}$	&18.1077 (13)&0.0000883 (57)&$9F_6-300F_o$&	12.42    &        $F_{93}$	&25.8693 (28)	&0.0000425 (57)	&$9F_6-294F_o$&6.28\\               
	$F_{45}$	&61.5845 (13)&0.0000872 (57)&		&14.55 &       
	$F_{93}$	&28.4648 (29)	&0.0000416 (57)	&$22F_o$&6.02\\                   
	$F_{46}$	&7.764931 (95)&0.0012720 (57)& $6F_o$&	159.33      &         $F_{95}$	&21.9901 (31)	&0.0000381 (57)&$9F_6-297F_o$&5.59\\               
	$F_{47}$	&54.7876 (14)&0.0000828 (57)&$7F_o+F_{21}(E)$ &	6.05  &         $F_{96}$	&108.8914 (19)	&0.0000615 (57)	&$4F_o+2F_{29}$&7.98\\             
	$F_{48}$	&56.8919 (13)&0.0000885 (57)&$4F_{45}-117F_o$&8.78        &      $F_{97}$	&104.1393 (23)	&0.0000524 (57)	&$2F_1$&6.12 \\                  
	$F_{49}$	&45.7828 (11)&0.0001020 (57)&$4F_{45}-155F_o$&10.97&            $F_{98}$	&103.1493 (30)	&0.0000398 (57)&$4F_o+2F_3$&5.49\\         
	\enddata
\end{deluxetable*}

\begin{figure}[ht!]
	\plotone{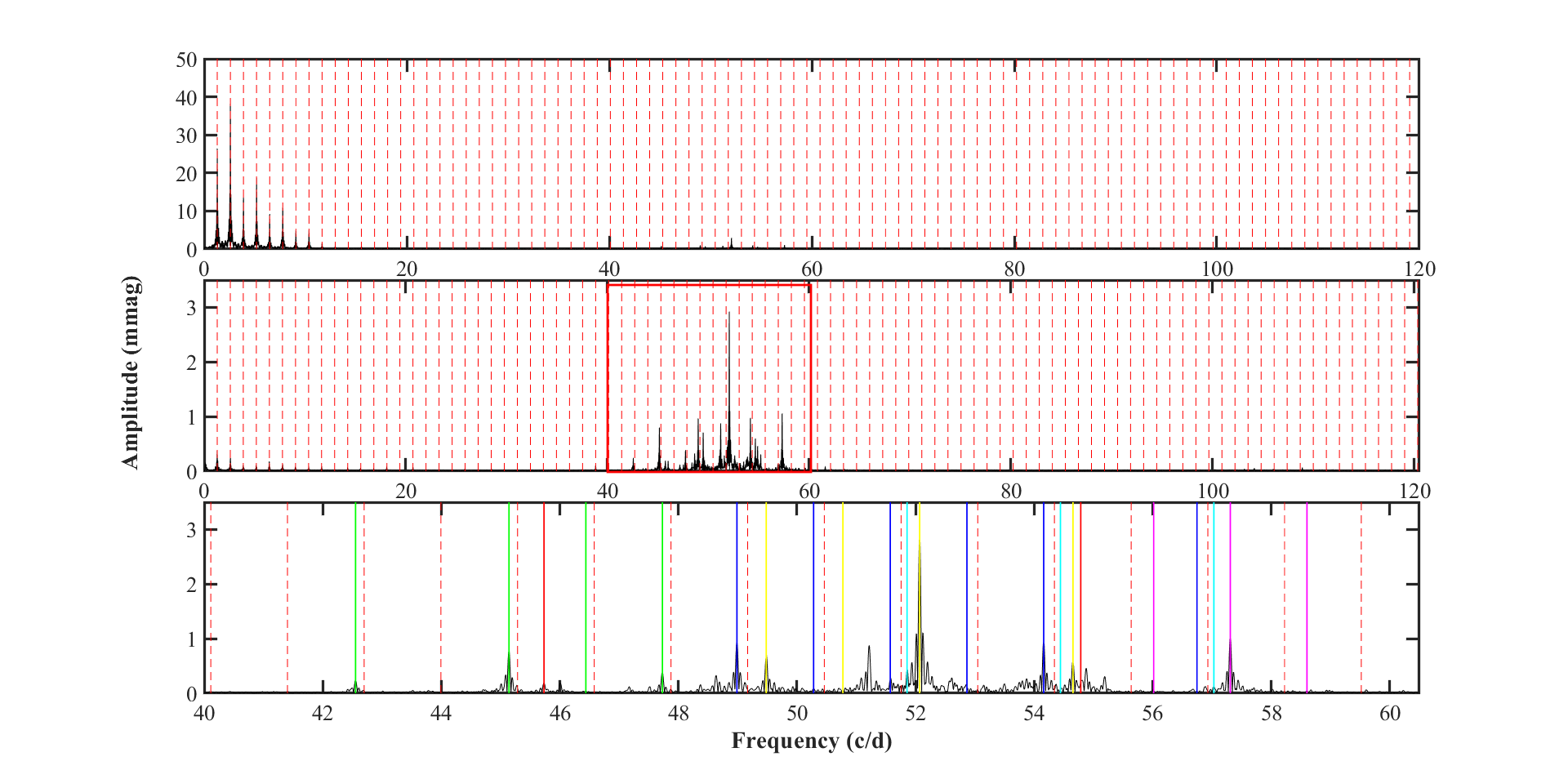}
	\caption{Fourier amplitude spectra of the TESS light curve for HZ Dra. Top panel: the original spectrum. Middle panel: the residual spectrum. Bottom panel: the same as the middle panel, but for a different view. The red dotted lines represent the harmonics of the orbital frequency. The coloured solid lines represent the six highest amplitude multiplets.\label{fig:figure 10}}
\end{figure}

At each step of the iteration, we selected the frequency with the highest amplitude and carried out a multi period least-square fit to the data using all the frequencies already detected. The data are then pre-whitened and the residuals are used for further analysis until an empirical signal-to-noise S/N=4.0 \citep{1993A&A...271..482B} threshold is reached. In the end, a total of 98 frequencies with a signal-to-noise S/N $>$ 4.0 were detected, which are listed in Table \ref{tab:table 5}. Based on the treatment reported by \cite{1999DSSN...13...28M}, the noise is calculated in the range of 1 d$^{-1}$ around each frequency. The uncertainties and amplitudes of these frequencies are also calculated. Figure \ref{fig:figure 10} shows Fourier amplitude spectra of the light residuals; the original spectrum is shown in the upper panel and the residual spectrum after pre-whitening is shown in the middle panel.

We checked the extracted frequencies and searched for possible orbital harmonics and liner combination frequencies in the form of $F_{k}$ = $F_{i}$ + $F_{\rm o}$ or $F_{k}$ = m $F_{i}$ + n $F_{j}$ \citep{2012AN....333.1053P,2015MNRAS.450.3015K} , where m and n are integers, $F_{i}$ and $F_{j}$ are the parent frequencies,  $F_{k}$ is the combination frequency, and $F_{\rm o}$=1.29375433 d$^{-1}$ is the orbital frequency of HZ Dra. If the amplitudes of both parent frequencies are greater than that of the presumed combination term, and the difference between the observed frequency and the predicted frequency is less than the frequency resolution $\delta f=1/\Delta T\approx 0.0028$ d$^{-1}$ \citep{1978Ap&SS..56..285L,2019AJ....157...17L}, where $\Delta T$ is the time range of the observations for sector 14-26, then the peak is accepted as a combination. As shown in Table \ref{tab:table 5}, we identified 78 linear combination frequencies, 4 orbital harmonic frequencies and 6 side lobe. Finally, 10 independent frequencies ($F_1$, $F_2$, $F_3$, $F_6$, $F_9$, $F_{13}$, $F_{21}$, $F_{29}$, $F_{32}$ and $F_{45}$) are detected.

\begin{figure*}[ht!]
	\plotone{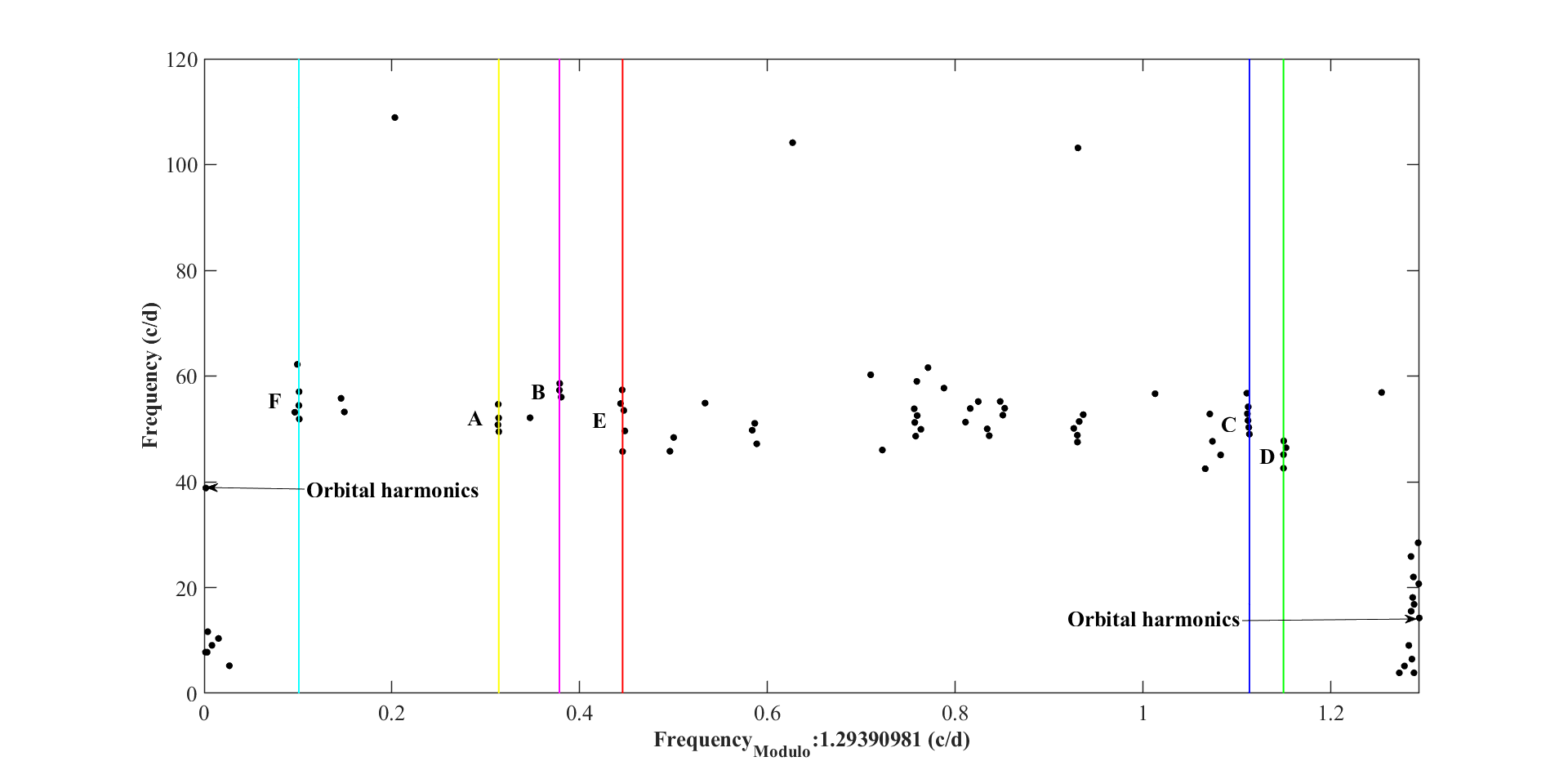}
	\caption{Echelle plot of all frequencies above 3 d$^{-1}$ versus they modulo the orbital frequency. The coloured solid lines are the same as those in the bottom panel of Fig.\ref{fig:figure 10}.\label{fig:figure 11}}
\end{figure*}

We also plot the Echelle diagram of all frequencies above 3 d$^{-1}$ modulo the orbital frequency $F_{\rm o}$ in Figure \ref{fig:figure 11}.
There are six possible pulsation frequencies, denoted by A ($F_1$, $F_7$, $F_8$, $F_{69}$), B ($F_2$, $F_{63}$, $F_{66}$), C ($F_3$, $F_4$, $F_{14}$, $F_{62}$, $F_{73}$, $F_{90}$), D ($F_6$, $F_{10}$, $F_{18}$, $F_{88}$), E ($F_{21}$, $F_{47}$) and F ($F_{29}$, $F_{30}$, $F_{36}$, $F_{89}$), are spaced as multiplets by the orbital frequency are thus discovered. They are six multiplets of the tidally split frequencies, similar to KIC 9851944 \citep{2016ApJ...826...69G}, KIC 6048106 \citep{2018MNRAS.474.5549S} and U Gru \citep{2019ApJ...883L..26B}. We suspect that  $F_9$, $F_{21}$, $F_{32}$ and $F_{45}$ may also be in a set of frequency multiplets here, but others with smaller amplitudes are not identified.

For studying further the pulsation modes of the independent frequencies, we calculated the pulsation constant $Q$ values for the 10 independent frequencies using the following basic relation for pulsating stars:
\begin{eqnarray}\label{equation (5)}
	Q=\mathit{P}_{\rm osc}\sqrt{\frac{\overline{\rho}}{\overline{\rho}_{\odot}} },
\end{eqnarray}
where the $\mathit{P}_{\rm osc}$ is the pulsation period and the mean density is $\overline{\rho}=0.3567 (\pm0.0061) \rho_\odot$. The $Q$ values for each frequency are used for comparison with the theoretical models \citep{1981ApJ...249..218F} for $M$=1.5 M$_{\odot}$. The probable $l$-degrees and the types of each frequency are listed in the Table \ref{tab:table 6}, except for $F_{45}$.
\begin{table*}
	\begin{center}
		\footnotesize
		\caption{Independent pulsating frequencies of HZ Dra. NR denotes non-radial pulsation.\label{tab:table 6}}
		\begin{tabular}{llllllc}\hline
			ID&Frequency (d$^{-1}$) &$Q\times 100$&$l$-degrees&Mode\\\hline
			$F_1$  & 52.069971 (42) &1.15 (1) &1& NRP6 \\
			$F_2$  & 57.31032 (11) &1.04 (1)&1& NRP7 \\
			$F_3$ & 48.98764 (12) & 1.22 (1)&3&NRP5\\
			$F_4$ & 45.14224 (15) & 1.33 (1) &1&NRP6 \\
			$F_9$ & 54.87759 (23) &1.09 (1)&0&R 6H \\
			$F_{13}$ & 45.787065 (3) &1.08 (1)&3&NRP7 \\
			$F_{21}$ & 55.19183 (41) &1.31 (1)&3&NRP5 \\
			$F_{29}$ & 51.85771 (88) &1.16 (1)&1&NRP6 \\
			$F_{32}$ & 7.766591 (92) &7.73 (7)&3&NRG2 \\
			$F_{45}$ & 61.5845 (13) &0.975 (9)&...... &...... \\
			\hline
		\end{tabular}
	\end{center}
\end{table*}
Our results show that 1 radial p-mode ($F_9$), 7 non-radial p-modes ($F_1$, $F_2$, $F_3$, $F_4$, $F_{13}$, $F_{21}$ and $F_{29}$) and 1 non-radial g-mode ($F_{32}$) are identified. Also, the dipole mode $F_1$ is really close to the frequency 52.17 d$^{-1}$ that had been found by \cite{2012MNRAS.422.1250L}. The $Q$ value for $F_{45}$ is less than 0.01, which is outside the range given by \cite{1981ApJ...249..218F}.

\section{DISCUSSIONS AND CONCLUSIONS} \label{sec:Conclusions}
The TESS light curve modelling using the W-D code shows that HZ Dra is a semi-detached eclipsing binary with the less massive secondary component almost filling its Roche lobe and a small mass ratio of 0.155 $\pm$ 0.001. The third light contribution is detected to be about 29$\%$. Combined with the previous radial velocity measure, fundamental parameters of the primary and secondary components are derived and listed in the lower part of Table \ref{tab:table 2}. In addition, we found that the variations in O$^{'}$Connell effect can be explained by an evolving hot and cool spot on the surface of the primary and secondary components, respectively, whose positions are roughly symmetric with respect to the inner Lagrangian point L1. According to \cite{2021AJ....161...46S}, the spots activities could be explained as the gravitational quadrupole coupling of the variation in the shape of a magnetically active secondary component \citep{1992ApJ...385..621A}. This leads the initial velocity of mass transfer from the secondary star to the primary one at the inner Lagrangian point L1 to be related to the magnetic activity. In addition to the gravitational quadrupole coupling, the magnetic activity of the cool spot regions could inhibit the energy transfer and prevent the regions from reaching thermal equilibrium, which could also lead to an expansion in these regions and a correlation between the magnetic activity and the initial velocity of mass transfer at the inner Lagrangian point L1. Therefore, it is observed that the positions of the cool and hot spots are roughly symmetrical with the internal Lagrangian point L1, the more intense the magnetic activity, the more intense the mass transfer can be achieved.

The cyclic variation of $O-C$ diagram for HZ Dra with a period of 1.05 years and semi-amplitude of 0.0033 days was firstly found in the present work. We tried to use the LTTE and Applegate mechanism to explain such cyclic variation of orbital period. As a result, our calculations reveal that the energy required for the Applegate mechanism is greater than the total radiative energy of the secondary star. We thus prefer the LTTE explanation via the presence of a close-in third body around the binary system. The minima mass of the tertiary star is calculated to be 1.61 $\pm$ 0.01 M$_{\odot}$ and its maximum semi-majox axis is estimated to be 0.92 $\pm$ 0.03 au. Assuming it is a main sequence star, we can estimate the luminosity of the third component using the following equations \citep{2000asqu.book.....C}.
\begin{eqnarray}\label{equation (6)}
	\log{M/M_{\odot}}=0.48-0.105 \mathit{M}_{\rm bol},
\end{eqnarray}
\begin{eqnarray}\label{equation (7)}
	\log(L/L_{\odot})=0.4\times(4.74-\mathit{M}_{\rm bol}),
\end{eqnarray}
The luminosity is therefore estimated to be about 5.81 L$_{\odot}$, which could contribute about 33 $\%$ of luminosity to the total system, and is in agreement with our photometric solution of 29$\%$ $\pm$ 4$\%$ within the error ranges.

According to the parallax $\pi$ = 5$^{''}$.4869 $\pm$ 0$^{''}$.0405 released by the Gaia mission \citep{2016A&A...595A...1G,2018A&A...616A...1G}, the corresponding Gaia distance of HZ Dra is calculated to be $d$ = 182.25 $\pm$ 7.38 pc. Combined with the semi-major axis of the third body, the maximum distance between the third body and the central binary is computed to be 1.49 $\pm$ 0.02 au. The corresponding angular separation can thus be calculated to be 0$^{''}$.0082 $\pm$ 0$^{''}$.0003, which is so small that we can just detect the luminosity contribution of the third body. However, we found no spectral component of the third body in our two low-resolution spectra. Medium- or high-resolution spectral observations may be needed in the future to further determine the presence of the tertiary star around HZ Dra.

Our multiple frequency analyses detect 1 radial p-mode, 7 non-radial p-modes and 1 non-radial g-mode. Meanwhile, we also identify a total of 6 multiplets, separated by the integral multiple of the orbital frequency. According to the theories of \cite{2003A&A...404.1051R, 2003A&A...409..677R}, these multiplets can be explained as a tidally split mode caused by the equilibrium tides of the close binary system with a circular orbit. These pulsation features show that the primary star of HZ Dra is a $\delta$ Scuti pulsating star, it pulsates in both p- and g-hybrid modes, and it is influenced by some physical process, such as tidal force, which makes its evolution different from that of a single $\delta$ Scuti pulsating star. 

The factors of a close-in third companion, spots activities and tidal split pulsations make HZ Dra an attractive oscillating eclipsing binary system of to study.

\begin{acknowledgments}
This work is supported by the International Cooperation Projects of the National Key R\&D Program (No. 2022YFE0127300), the National Natural Science Foundation of China (No. 11933008), the Young Talent Project of  ``Yunnan Revitalization Talent Support Program" in Yunnan Province, the basic research project of Yunnan Province (Grant No. 202201AT070092), CAS ``Light of West China" Program and the Natural Science Foundation of Anhui Province (2208085QA23). This work has made use of data from the European Space Agency (ESA) mission Gaia. (\href{https://www.cosmos.esa.int/gaia}{https: //www.cosmos.esa.int/gaia}), processed by the Gaia Data Processing and Analysis Consortium (DPAC; \href{https://www.cosmos.esa.int/web /gaia/dpac/consortium}{https://www.cosmos.esa.int/web /gaia/dpac/consortium}). Funding for the DPAC has been provided by national institutions, in particular the institutions participating in the $Gaia$ Multilateral Agreement. The $TESS$ data presented in this paper were obtained from the Mikulski Archive for Space Telescopes (MAST) at the Space Telescope Science Institute (STScI). STScI is operated by the Association of Universities for Research in Astronomy, Inc. Support to MAST for these data is provided by the NASA Office of Space Science. Funding for the $TESS$ mission is provided by the NASA Explorer Program. We thank the anonymous referee for valuable comments and suggestions, which have improved the manuscript greatly.
\end{acknowledgments}

\bibliography{Ref}{}
\bibliographystyle{aasjournal}



\end{document}